\documentclass[12pt,reqno]{amsart}
\usepackage{graphicx}
\usepackage{amssymb,amscd,amsbsy}
\usepackage{amsthm}
\newcommand{\noi}{\noindent}

\newcommand{\lab}{{\boldsymbol \lambda}}
\newcommand{\sigb}{{\boldsymbol \sigma}}
\newcommand{\gamb}{{\boldsymbol \gamma}}

\newcommand{\Z}{{\mathbb Z}}
\newcommand{\C}{{\mathbb C}}
\newcommand{\RB}{{\mathbb R}}

\newcommand{\CP}{{\mathbb C}{\mathbb  P}}

\newcommand\Dg{\mathfrak D}

\newcommand{\HH}{{\mathcal H}}
\newcommand{\MM}{{\mathcal M}}

\newcommand{\X}{\mathcal X}

\def\O{\Omega}

\def\F{{\mathcal W}}
\def\P{\mathcal X}
\def\Pi{\mathcal D}
\def\[{\left[}
\def\]{\right]}
\def\({\left(}
\def\){\right)}

\def\l{\lambda}

\def\m{\mu}
\def\lt{\frac{\lambda}{2}}
\def\psib{\overline{\psi}}
\def\D{\Delta}
\def\d{\delta}

\def\ea{\epsilon_a}
\def\epm{\epsilon_{\pm}}

\def\gh{\hat{\boldsymbol \gamma}}
\def\Ph{\hat{\Psi}}

\def\G{\Gamma}
\def\Qh{\hat{Q}}
\def\Ph{\hat{P}}
\def\Gh{\hat{\Gamma}}
\def\12{{1\over 2}}

\newcommand{\ee}{{\mathfrak e}}

\newcommand{\e}{{\boldsymbol e}}
\newcommand{\kk}{{\boldsymbol k}}
\newcommand{\f}{{\boldsymbol f}}

\newcommand{\LLL}{{\boldsymbol L}}

\newcommand{\gb}{{\boldsymbol g}}

\newcommand{\p}{\partial}
\newcommand{\eeq}{\end{equation}}
\newcommand{\beq}{\begin{equation}}
\newcommand{\bay}{\begin{eqnarray}}
\newcommand{\ey}{\end{eqnarray}}
\newcommand{\bey}{\begin{eqnarray*}}
\newcommand{\eey}{\end{eqnarray*}}

\newcommand{\R}{\operatorname{res}}

\newtheorem{thm}{\hspace{\parindent}Theorem}[section]

\newtheorem{lem}[thm]{\hspace{\parindent}Lemma}

\newtheorem{exa}[thm]{Example}

\theoremstyle{remark}
\newtheorem{rem}[thm]{Remark}
\newtheorem*{rem*}{Remark}

\begin{document}

\newcommand{\vse}{\vspace{.2in}}
\numberwithin{equation}{section}

\title{\bf the atiyah-hitchin bracket for the cubic nonlinear schr\'{o}dinger equation. ii. Periodic potentials.}
\author{K.L.  Vaninsky}
\thanks{ The work is partially supported by NSF grant DMS-9971834.}
\begin{abstract}
This is the second in a series of  papers on Poisson formalism  for the cubic nonlinear Schr\"{o}dinger 
equation with repulsive nonlinearity.  In this paper we consider periodic potentials.   
The  inverse spectral problem for the periodic auxiliary Dirac operator leads to  a  
hyperelliptic Riemann surface $\G$. Using the spectral problem we introduce  on this Riemann surface a meromorphic  function $\P$. 
We call it the Weyl function, since it is closely related to the classical Weyl function discussed in the first paper.   
We show that  the  pair $(\G,\P)$ carries a natural Poisson structure. We call it the deformed Atiyah--Hitchin  bracket.  The Poisson bracket on the phase space is the  image of the deformed Atiyah--Hitchin bracket 
under the inverse spectral transform.

\end{abstract}
\maketitle

\setcounter{section}{0}
\setcounter{equation}{0}
\section{Introduction.} 
\subsection{\bf Statement of the problem.}
The fact that  equations integrable by the method of the inverse spectral transform are Hamiltonian systems was  realized  in the very 
early days of the theory.  Gardner, Zakharov and Faddeev \cite{G,ZF},  found  that the Korteveg de Vries equation 
\footnote{Prime $'$  signifies the derivative in the   variable $x$ and dot $\bullet$ the derivative with 
respect to time.}  
$$
u^{\bullet}= 6u'u -u''',\qquad\qquad\qquad\qquad u=u(x,t);
$$ 
on the line with 
rapidly decaying initial data  can be written as a Hamiltonian system
$$
u^{\bullet}=\{u,\HH\},
$$ 
with  Hamiltonian
$
\HH=\int  u^3 + \frac{1}{2} (u')^2 \, dx
$
and the bracket
\beq\label{gfz}
\{A,B\}=\int \frac{\d A}{\d u(x)} \frac{\partial}{\partial x} \frac{\d B}{\d u(x)}\, dx.
\eeq  
 
Soon after   Zakharov and Manakov, \cite{ZM}, integrated  the nonlinear Schr\"{o}dinger equation with repulsive nonlinearity
$$
i \psi^{\bullet}= -\psi'' + 2 |\psi |^2 \psi,  
$$
where $\psi(x,t)$ is  a  smooth complex function on the line with rapid decay at infinity. 
The equation is a Hamiltonian system
$$
\psi^{\bullet}= \{\psi, \HH\},
$$
with  Hamiltonian
$\HH=\frac{1}{ 2}\int   |\psi'|^2 +|\psi|^4 \, dx=${\it energy} and the classical bracket
\beq\label{pb}
\{A,B\}=2i \int \frac{\d A}{ \d \psib(x)}\frac{\d B}{ \d
\psi(x)}- \frac{\d A}{ \d \psi(x)}\frac{\d B}{ \d \psib(x)}\, dx.
\eeq
The NLS equation will surve as our model example. The KdV case  is  more subtle and will be considered in \cite{V5}. 

It was demonstrated by Novikov,  Dubrovin, Its--Matveev and McKean--Moerbecke, \cite{DMN,MM}, that the periodic problem for the KdV equation is connected with  hyperelliptic Riemann surfaces. The periodic problem for the cubic NLS is also connected with  hyperelliptic Riemann surfaces, \cite{I}.

At the present time we know  numerous  examples of integrable systems and various Hamiltonian formulations of them.   
Until now it was not known how to  obtaining the {\it Poisson formalism} from the 
corresponding Riemann surface. The goal of the present paper is to make a new step in this direction. 
Namely we relate bracket  \ref{pb} with the Poisson structure on the meromorphic functions defined on the  corresponding hyperelliptic Riemann surfaces. This 
Poisson bracket we call the deformed Atiyah--Hitchin bracket. 

In our previous paper \cite{V1}  we introduced  the spectral cover for a class of general  potentials. This  is two sheeted covering of the 
complex plane cut along the  real line. The spectral cover is an open  manifold which consist of four (disconnected) copies of the complex half--plane. On this cover we defined the classical Weyl function. We showed that the Weyl function carries the  Atiyah--Hitchin bracket. 
The formula for the bracket  has a different shape for  different parts of the cover. 

The construction of the present  paper is  a compactification of the spectral cover. In the case of periodic finite gap potentials the spectral 
cover can be glued into a plane curve biholomorphicaly equivalent to a compact smooth hyperelliptic Riemann surface. 
The Weyl function can be  analytically extended across the glued edges.   Now the spectral cover  is simply connected  and the Poisson 
bracket on Weyl functions  is described by a single formula. This is  the deformed Atiyah--Hitchin bracket. We demonstrate that the 
Poisson bracket on the phase space is the  image of this deformed AH bracket under the inverse  spectral transform.

We would like to make one  historical remark. 
Novikov and Veselov in their pioneering paper \cite{NV} singled out a class of brackets for the KdV equation. They call these brackets {\it analytic brackets compatible with  algebraic geometry}. The brackets  can be written in terms of singularities of the Floquet solutions or, in other words,  eigenfunctions of the auxiliary spectral problem with  special monodromy properties.  The Gardner bracket \ref{gfz}, the Lenard-Magri bracket \cite{MA}, {\it etc.}, are examples of such brackets.  Similar situation was noted for the NLS hierarchy, \cite{DN}.

The approach of this paper is conceptually different from the  approach of Novikov {\it at all}. We consider the bracket where the corresponding  functions are holomorphic, while 
Novikov {\it at all} express the brackets in terms of singularities. Nevertheless the term {\it analytic brackets compatible with  algebraic geometry} 
coined by them describes what happens here in the most precise way. 
 
In the rest of the introduction we describe our strategy and  results. We try to sweep all technicalities under the rug in order to give a reader a clear   geometrical picture. 
 
\subsection{\bf The Atiyah-Hitchin bracket.}
Atiyah and Hitchin,   \cite{AH},  introduced a symplectic structure on the space of meromorphic    maps   
$\P(\l): \CP^1 \rightarrow \CP^1$ of the form 
$$
\P(\l)=-\frac{q(\l)}{p(\l)},
$$
where $q(\l)$ is a polynomial of degree $N-1$ and $p(\l)$ is a monic polynomial of degree $N$ with distinct roots.  
The  parameters $\l_1,\hdots,\l_{N}, q(\l_1),\hdots,q(\l_{N}),$ are complex coordinates on this space and   
 $\d$ denotes a variation of these coordinates. The Atiyah--Hitchin   nondegenerate close 2-form  
$\omega$ is defined by the formula
$$
\omega=\sum_{k=1}^{N} \frac{\d q(\l_k)}{q(\l_k)} \wedge \d \l_k.
$$

The corresponding Poisson bracket is specified  by canonical relations:
$$
\{q(\l_n), \l_k\}=\d^n_k \; q(\l_n);\qquad \quad
\{\l_n, \l_k\}=\{q(\l_n), q(\l_k)\}=0.
$$
The bracket turns  the space of maps $\P(\l): \CP^1 \rightarrow \CP^1$  into a Poisson manifold. 
Consider $\P(\l)$ and $\P(\m)$ where  the variables  $\l$ and $\m$ are fixed  and away from the poles. Then  $\P(\l)$ and $\P(\m)$ (considered as  functions of $\P$) are  
functions of coordinates  $\l_1,\hdots,\l_{N},$ $ q(\l_1),\hdots,q(\l_{N})$. 
As it was  demonstrated by Faybusovich and Gekhtman, \cite{FG}, the bracket for $\P(\l)$ and $\P(\m)$,  is given by the formula
\beq\label{AH}
\{\P(\l),\P(\m)\}=\frac {(\P(\l)-\P(\m))^2}{\l-\m}.
\eeq
This  formula   is much more general than its coordinate version, see \cite{V1}.

In this  paper we construct   Poisson structure on  the space of pairs $(\G,\P)$ where $\Gamma$ 
is  a  hyperelliptic Riemann surface of infinite genus associated with 
an inverse spectral problem for the  Dirac operator  
\beq\label{dirac}
\Dg \f= \[\(\begin{array}{ccccc} 1& 0 \\ 0 &  -1  \end{array}\) i\partial_x +
\(\begin{array}{ccccc}   0& -i \psib \\
 i \psi & 0 \end{array}\)\] \f= \lt \f,\qquad\qquad \f=\[\begin{array}{ccccc} f_1 \\f_2 \end{array}\]
\eeq
in the class of all smooth periodic  potentials $\psi(x+2l)=\psi(x)$  and $\P: \G \rightarrow \CP^1$ is the Weyl  function.   
 An isospectral deformations  of the Dirac operator (which preserve $\G$) are   the flows of cubic nonlinear Schr\"{o}dinger   hierarchy.
We construct  the Poisson structure on  $(\G,\P)$   
using the   Poisson bracket \ref{pb}.  

We illustrate our strategy using   the Camassa--Holm equation and 
the open Toda lattice. In both cases the Riemann surface  is reducible with components being  copies of $\CP^1$. Each component of such curve has a global uniformization parameter.
These curves with nodal singularities appear in the  compactification\footnote{ The situation is similar 
to the Deligne-Mumford compactification of the space of smooth curves by stable curves, \cite{DM}.} of the space of smooth hyperelliptic  curves (possibly of infinite genus). 
   This is a great simplification of the  hyperelliptic case 
  since each component  can be treated by analysis methods. Such reducible curves were first considered in the theory of completely integrable systems 
by McKean in the beginning of 1980's, \cite{MC}. The Baker-Akhiezer function for such curves was introduced only recently by Krichever 
and the author, \cite{KV}.   We describe the Poisson brackets  for the Weyl functions defined on  these boundary curves.

\begin{exa} {\bf The Camassa--Holm equation.} \end{exa} 
The simplest situation  when  formula \ref{AH} appears is  
the Camassa--Holm equation 
$$
\frac{\p v}{ \p t} + v\frac{\p v}{ \p x} +\frac{\p  }{ \p x}G\[v^2 +\frac{1}{ 2} \(\frac{\p v}{ \p x}\)^2\]=0$$
in which $t\geq 0$ and $  -\infty < x < \infty$, $v=v(x,t)$ is velocity, and $G$ is inverse to $1-d^2/dx^2$ {\it i.e.,}
$$
G: f(x)\rightarrow \frac{1}{ 2} \int\limits_{-\infty}^{+\infty} e^{-|x-y|} f(y) dy.
$$
The CH equation can be formulated as a Hamiltonian system 
$$
m^{\bullet} +\{m,\HH\}=0,\quad\qquad\qquad\qquad m=v-\frac{\p^2v}{\p x^2};
$$
with the Hamiltonian
$$
\HH=\frac{1}{ 2}\int_{-\infty}^{+\infty} mv\, dx=\text{energy}
$$
and the bracket
\beq\label{chpb}
\{A,B\}=\int_{-\infty}^{+\infty}{\delta A\over \delta m} \(mD+Dm\) {\delta B\over \delta m}\, dx.
\eeq

The  CH equation preserves the Dirichlet spectrum of the  string spectral problem, 
$$
\frac {\p^2 f(\xi)}{\p \xi^2}+ \l g(\xi)f(\xi)=0,\qquad\qquad\qquad\qquad -2\leq \xi \leq 2.
$$
The variables $\xi$ and $x$ are related by
$$
x\longrightarrow \xi =2\tanh {x\over 2}.
$$
Also the potential $g(\xi)$ is related to $m(x)$ by the formula $g(\xi)=m(x)\cosh^4{x\over 2}$.

The Riemann surface $\Gamma$  associated   with the 
spectral problem consists of two   components $\Gamma_-$ and $\Gamma_+$  (see Figure 1), two copies of  $\CP^1$. The components are glued to each other  at the points of the Dirichlet spectrum $\l_k,\;k=1,2,\hdots$. A point on the curve is  denoted by $Q=(\l,\pm)$, where $\l$ is the spectral parameter and the sign $\pm$ refers to the  component.   An infinities of $\Gamma_{\pm}$ are  denoted by $P_{\pm}$ correspondingly.

\begin{figure}[ht]
\includegraphics[width=0.80\textwidth]{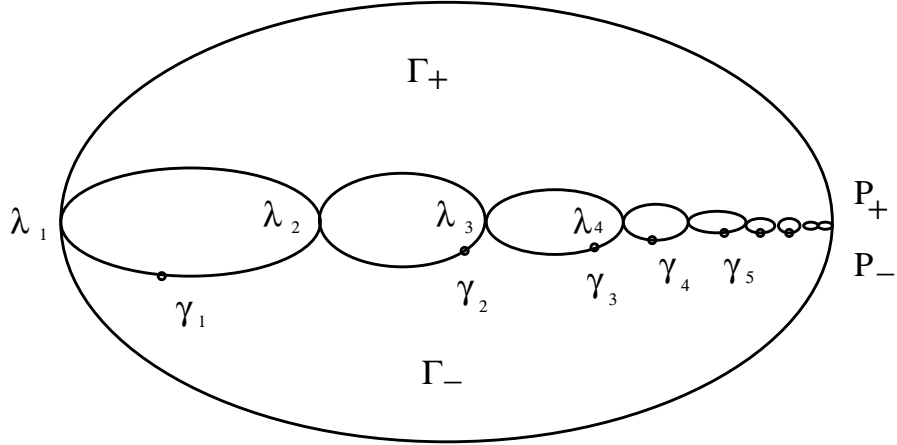}
\caption{The  reducible Riemann surface with infinitely many nodal points for the CH equation.}
\end{figure}

The Baker--Akhiezer function $e(\xi,Q)$ is a function  on $\Gamma$ which  depends on the variable $\xi, -2\leq \xi \leq +2$, as a parameter. In this 
parameter the BA function  is a solution of the string spectral problem with $\l=\l(Q)$. 
The BA function  is  holomorphic outside simple poles at the points of the divisor $\gamb_k=(\mu_k,-),\; k=1,2,\hdots,$ and   singularities  at infinities $P_{\pm}$.
The Baker--Akhiezer function    satisfies the gluing condition 
$$
e(\xi,(\l_k,+))=e(\xi,(\l_k,-)),\qquad\qquad\qquad k=1,2,\hdots. 
$$ 

For  $Q\in\Gamma _-$ we define  the Weyl function by the formula 
$$
\P(\l)=\frac{\partial}{\partial \xi} \log  e(\xi,Q)|_{\xi=-2},\qquad\qquad\qquad \l=\l(Q).
$$ 
The function $\P$ is a real meromorphic map $\CP^1 \rightarrow \CP^1$.  
After a change of the  spectral parameter $\l \rightarrow -{1/\l}$  the function $\P$ takes the form 
$$
\P(\l)=-\frac{1}{4}+\sum_{k=1}^{\infty} \frac{\rho_k}{\l_k-\l}.
$$
The Poisson bracket \ref{chpb} computed for $\P$  is given by  formula \ref{AH}.
  We refer to \cite{V3} for details.

\begin{exa} {\bf The open Toda hierarchy.} \end{exa}
This   $N$--particles  system is another example where  the space of maps $\CP^1\rightarrow \CP^1$ occurs.
The Hamiltonian of the system is 
$$
H=\sum\limits_{k=0}^{N-1}\frac{p_k^2}{ 2} +\sum\limits_{k=0}^{N-2}e^{q_k-q_{k+1}}.
$$
Introducing the classical Poisson bracket
\beq\label{tpb}
\{f,g\}=\sum\limits_{k=0}^{N-1} \frac{\p f}{\p q_k} \frac{\p g}{ \p p_k}- 
                                \frac{\p f}{ \p p_k} \frac{\p g}{\p q_k},  
\eeq
we write the equations of motion as  
\begin{eqnarray}
q_k^{\bullet}&=&\{q_k,H\}= p_k, \nonumber\\
p_k^{\bullet}&=&\{p_k,H\}=-e^{q_k-q_{k+1}}+e^{q_{k-1}-q_{k}},\qquad\qquad\qquad
k=0,\ldots,N-1. \nonumber
\end{eqnarray}
We put $q_{-1}=-\infty,\;  q_N=\infty$ in all formulas.

The Toda flow preserves the spectrum $\l_1<\hdots < \l_N$  
 of the three diagonal Jacobi matrix 
$$
L=\left[\begin{array}{ccccc}
v_0      & c_0       & 0         &\cdots        & 0\\
c_0      & v_1       & c_1       &\cdots        & 0\\
\cdot        &  \cdot    & \cdot     &\cdot         & \cdot\\
0    &  \cdots         & c_{N-3}         & v_{N-2}     & c_{N-2}\\
0       & \cdots    & 0         & c_{N-2}      & v_{N-1}
\end{array}\right],
$$
where 
$$
c_k=e^{q_k-q_{k+1}/2},\quad\qquad\qquad v_k=-p_k.
$$

The Riemann surface $\Gamma$  associated   with this 
spectral problem consists of two   components $\Gamma_-$ and $\Gamma_+$, two copies of $\CP^1$. The components are glued to each other   
at the points of the  spectrum $\l_k,\;k=1,\hdots, N$; (see Figure 2).

\begin{figure}[ht]
\includegraphics[width=0.80\textwidth]{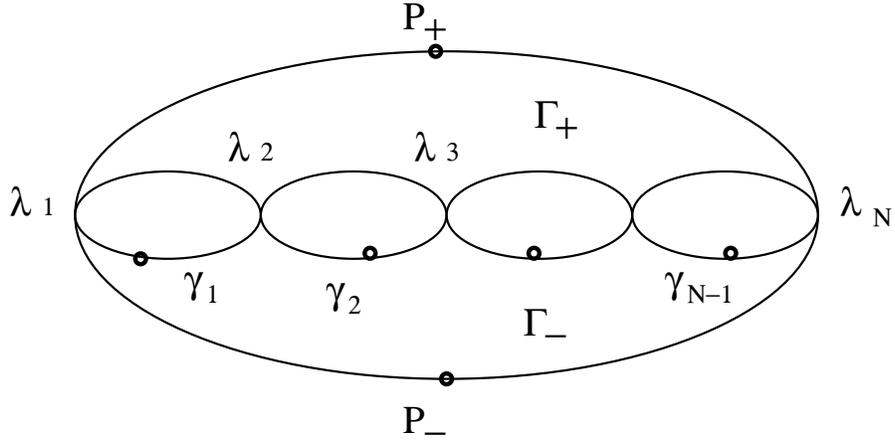}
\caption{The  reducible Riemann surface with finite number of nodal points for the open Toda.}
\end{figure}

The Baker--Akhiezer function $e(n,Q)$ is a function  on $\Gamma$ which  depends on the variable $n,\, n=1,\hdots, N$, as a parameter. In this 
parameter it satisfies the three term reccurence relation produced by the matrix $L$. 
The BA function is  holomorphic outside simple poles at the points of the divisor $\gamb_k=(\mu_k,-),\; k=1,\hdots,N-1;$ and  multiple poles   at infinities $P_{\pm}$.
The Baker--Akhiezer function    satisfies the gluing condition 
$$
e(n,(\l_k,+))=e(n,(\l_k,-)),\qquad\qquad\qquad k=1,\hdots,N. 
$$ 

When $Q\in\Gamma _-$ we define  the Weyl function by the formula 
$$
\P(\l)=\frac{1}{c_0e(1,Q)- \l+ v_0},\qquad\qquad\qquad \l=\l(Q).
$$
 The function $\P$ is a rational map $\CP^1 \rightarrow \CP^1$. It  can be expanded in simple fractions  
$$
\P(\l)= \frac{\rho_1}{\l_1-\l}+\hdots + \frac{\rho_{N}}{\l_{N}-\l},
$$
with  $\l_k$ being  
real and $\rho_k >0 $. Moreover, the  spectral problem imposes an additional condition 
$
\sum \rho_k=1
$
or equivalently
\beq\label{rrf}
\R_{P_-}\P(\l) d\l=1.
\eeq
The Poisson bracket \ref{tpb} computed for $\P$ takes the form
\beq\label{MAH}
\{\P(\l),\P(\m)\}=(\P(\l)-\P(\m))\(\frac {\P(\l)-\P(\m)}{\l-\m}-\P(\l)\P(\m)\) .
\eeq
This bracket is  a Dirac restriction of the AH bracket  to  the submanifold \ref{rrf}. The bracket \ref{MAH} is degenerate with the 
Casimir
$$
\sum \l_k=-\sum p_k.
$$ 
 For the unrestricted AH bracket   this quantity is canonically conjugate to $\sum \rho_k$,   see \cite{V2} for details.

\subsection{\bf  The   deformed Atiyah-Hitchin bracket.}

These examples illustrate  the following general   scheme. We associate to  every point of the phase space $\MM$ of an integrable system a set of algebraic-geometrical data. 
These data are the  Riemann surface $\G$ and the  meromorphic function $\P:\G\rightarrow \CP^1$,
$$
\MM \longrightarrow (\G,\P).
$$
We call this map the   direct spectral transform.  

For a generic periodic potential  of the Dirac   spectral problem \ref{dirac} one has to consider curves of  infinite genus. For the so--called 
finite gap potentials we  describe the finite genus curves in the image of the   direct spectral transform in purely geometrical terms. Generic 
infinite gap case has little to add to this picture.

The nonsingular curve \footnote{The curve we are considering  here is a resolution of singularities 
for some plane curve with infinitely many intersection points.} $\G$ of genus $g$ has a meromorphic function $\l(Q),\, Q \in \G$ of degree two 
with two simple poles at two points $P_-$ and $P_+$ which we call infinities. 
Thus the  curve is hyperelliptic. 
Namely, there are $2g+2$ 
critical points $\lab^\pm_k,\;\; k=1,\hdots,g+1,$ of the function $\l(Q)$.  The  critical values $\l^\pm_k=\l(\lab^\pm_k)$  are  points of the simple periodic /antiperiodic spectrum of the Dirac operator and branch points of this surface. The curve is  defined by 
$$
\G=\{(\l,R)\in \C^2:\; R^2=-\prod_{k=1}^{2g+2}(\l-\l_k)\}.
$$

The  self-adjoint Dirac spectral problem   requires  the branch points of $\G$ to be  real.  
The hyperelliptic Riemann surface $\G$ has a standard  anti--holomorphic involution $\ea:\G \rightarrow \G$ permuting infinities. 
The  branch points are fixed points of this involution :
$$
\ea \lab^\pm_k =\lab^\pm_k.
$$
The branch points lie on the fixed real ovals of the anti--holomorphic involution. There are $g+1$  real ovals $a_1,\hdots,a_{g+1}$ (see Figure 3).

\begin{figure}[ht]
\includegraphics[width=0.80\textwidth]{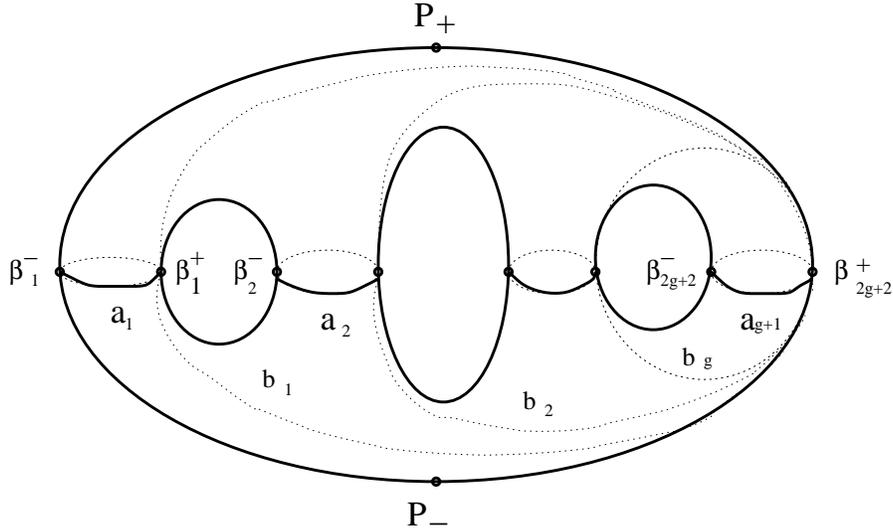}
\caption{The  hyperelliptic Riemann surface.}
\end{figure}

The   spectral problem imposes another  restriction on the corresponding hyperellitic Riemann surfaces. The Riemann surface 
caries a Floquet muliplier $w(Q)$. The transcendental function $w(Q)$ is holomorphic in the finite part of the curve and
$w(Q)\sim e^{\pm i l \l(Q)}$ for $  Q$   near $P_{\pm}$. 
At the branch points 
$
w(\lab^\pm_k)=\pm 1.
$
The function $w(Q)$ takes  values $\pm 1$ also at the points of the double spectrum.

Now when we are back to the generic potentials we describe the function $\P$ which we call the Weyl function on the Riemann surface. The spectral problem \ref{dirac} 
determines the $2\times 2$ monodromy matrix for the shift 
$x \longrightarrow x+2l$ over the period of the potential. The eigenvector of the monodromy matrix trivializes on the spectral curve $\G$. The eigenvector is defined up to a multiplicative constant, but the ratio of its components is defined uniquely. This ratio is the meromorphic function 
$\P(x,Q),\; Q\in \G,$ and can be expressed in terms of the Baker-Akhiezer function.  The ratio plays the role of the logarithmic derivative that we used for the CH  and Toda lattice.

The direct  spectral transform  maps the Poisson bracket \ref{pb} defined on the phase space $\MM$ of all smooth $2l$-periodic potentials    to the space of pairs $(\G, \P)$. 
 This bracket on the target space $(\G, \P)$ we call the deformed AH bracket.  To state the formula for it   we introduce the deformation factor 
$$
\O(Q)=\frac{w^2(Q)+1}{w^2(Q)-1}.
$$
At the branch or intersection points  the function $\O$ has  
poles because  $w^2(Q)= 1$.
The deformed AH bracket  for two functions $\P(Q)=\P(x,Q)$ and  $\P(P)=\P(x,P)$   is given by the formula
$$
\{\P(Q),\P(P)\}=-2\times \frac{\(\P(Q)-\P(P)\)^2}{\l-\mu}\times \frac{\O(Q)+\O(P)}{2}.
$$ 
Or  course, we assume that  $Q$ and $P$ are  not at the  poles of the function $\O$ or $\P$.

The formula   differs  from   the ''pure" AH bracket \ref{AH}  by multiplication   
on  a sum of deformation factors at two points $P$ and $Q$.  
The origin of the deformation factor lies in the fact that $\G$ covers the plane of the spectral parameter $2:1$. At the branch or intersection 
points the function $\l(Q)$ can not be taken as a local parameter  and the deformation factor  remedies  the situation.

The direct spectral transform 
can be inverted. 
The inverse   map 
$$
(\G,\P) \longrightarrow \MM
$$
is called the inverse spectral transform. The image of the deformed AH on the phase space under the inverse  spectral transform  is the 
standard Poisson bracket \ref{pb}. Therefore, the deformed AH bracket can be taken as a starting formula for construction of the Poisson formalism\footnote{The same formula produces the Gardner--Zakharov--Faddeev bracket \ref{gfz} for the KdV equation, \cite{V5}.}.

\subsection{Content of the paper.} The paper is divided into two parts. In Section 2 we discuss the properties of the direct and 
inverse spectral transform. In Section 3 we compute the image of the bracket for the direct and inverse spectral transform. We also discuss the construction of the canonical variables.

Finally, 
the author  would like to thank I. Krichever, S. Novikov, S. Natanzon, H. McKean and M. Shapiro for stimulating discussions.

\section{The  Spectral Problem.}

\subsection{The NLS hierarchy.}
The  NLS equation
\bay\label{nls}
i \psi^{\bullet}= -\psi'' + 2 |\psi |^2 \psi, 
\ey
where $\psi(x,t)$ is  a  smooth $2l$--periodic, $\psi(x+2l)=\psi(x)$, complex function   
is  a Hamiltonian system
$$
\psi^{\bullet}= \{\psi, \HH\},
$$
with  Hamiltonian
$\HH=\frac{1}{ 2} \int_{-l}^{l} |\psi'|^2 +|\psi|^4 \, dx=${\it energy} and the bracket
\beq\label{nlspb}
\{A,B\}=2i \int \frac{\d A}{ \d \psib(x)}\frac{\d B}{ \d
\psi(x)}- \frac{\d A}{ \d \psi(x)}\frac{\d B}{ \d \psib(x)}\, dx.
\eeq
The  NLS Hamiltonian $\HH=\HH_3$ is one in the infinite series of commuting  integrals of motion
\bey
\HH_1&=&\frac{1}{ 2} \int_{-l}^{l} |\psi|^2 dx,\\
\HH_2&=&\frac{1}{ 2i} \int_{-l}^{l} \psi \bar{\psi}'  dx,\\
\HH_3&=&\frac{1}{ 2} \int_{-l}^{l} |\psi'|^2 +|\psi|^4 \, dx,\quad etc.\\
\eey
Hamiltonians produce an infinite hierarchy of  flows $e^{tX_m},\; m=1,2,\ldots$. 

The NLS equation \ref{nls}
is a compatibility condition for the commutator
\bay\label{com}
[\partial_t- V_3,  \partial_x-V_2]=0, 
\ey
with\footnote{
Here and below  $\sigma$  denotes the {\it Pauli matrices}
$$
\sigma_1=\left(\begin{array}{ccccc}
 0& 1\\
1 & 0
\end{array}  \right),  \nonumber \quad
\sigma_2=\left(\begin{array}{ccccc}
0 & -i\\
i &  0
\end{array} \right), \nonumber \quad
\sigma_3=\left(\begin{array}{ccccc}
1 & 0\\
0 & -1
\end{array} \right). \nonumber
$$
}
$$
V_2 = - \frac{i \l }{ 2} \sigma_3 +Y_0 =
\(\begin{array}{ccc}
 - \frac{i\l}{ 2} &  0\\
 0 & \frac{i\l}{ 2} \end{array}\)  +
\(\begin{array}{ccccc}
  0& \overline{\psi} \\
 \psi & 0 \end{array}\)
$$
and
$$
V_3 = \frac{\l^2}{ 2}i \sigma_3 -\l Y_0 + |\psi|^2 i\sigma_3 -i \sigma_3 Y_0'.
$$
We often omit the lower index and write $V=V_2$. Each flow $e^{tX_m}$ of the hierarchy can be written in the form \ref{com}
with the suitable operator $\partial_t- V_m$.

\subsection{\bf The direct spectral problem for the Dirac operator.} We cite here freely the results of \cite{MV}.
We assume that the periodic potential $\psi(x)$ is defined on the entire line. The commutator relation produces   the auxiliary linear problem  
\beq\label{spro}
\f'(x,\l)=V \f(x,\l), \qquad\qquad\qquad \f(x,\l)=\[  \begin{array}{ccccc}
                 f_1(x,\l)\\
                 f_2(x,\l) \end{array}\].  
\eeq
This can be written as an eigenvalue problem for the Dirac operator
\beq\label{dirc}
{\Dg} \f= \[\(\begin{array}{ccccc} 1& 0 \\ 0 &  -1  \end{array}\) i\partial_x +
\(\begin{array}{ccccc}
 0& -i \overline{\psi} \\
 i \psi & 0 \end{array}\)\] \f= \lt \f.
\eeq

We introduce a  $2\times 2$  transition  matrix $M(x,y,\l)$: 
$$
M(x,y,\l)= \[  \begin{array}{ccccc}
                 m_{11} &  m_{12}\\
                m_{21} &  m_{22} \end{array}\] (x,y,\l),
$$
which is a  solution of  the equation
$$
M'(x,y,\l)=V M(x,y,\l),\qquad\qquad\qquad M(y,y,\l)=I.
$$

Let us define  the monodromy matrix
$$
M(x,\l)=M(x+2l,x,\l)= \[  \begin{array}{ccccc}
                 M_{11} &  M_{12}\\
                M_{21} &  M_{22} \end{array}\] (x,\l).
$$
We have obvious relation 
$$
M(x,\l)=M(x+2l,l,\l)M(-l,\l)M^{-1}(x+2l,l,\l).
$$
Therefore, all  matrices $M(x,\l)$ are   similar to  $M(-l,\l)$.
The monodromy  matrix $M(-l,\l)$ is unimodular  because $V$ is traceless.
The quantity  $\Delta(\l)=\frac{1}{ 2} \text{trace}  M(-l,\l)$  is called a discriminant. The symmetry of the matrix $V(x,\l)$
$$
\sigma_1 \overline{V}(x,\l)\sigma_1= V(x,\bar{\l})
$$
produces  the same relation for the transition  matrix 
\bay\label{real}
\sigma_1 \overline{M}(x,y,\l)\sigma_1= M(x,y,\bar{\l}).
\ey
This implies in particular $\overline{\Delta}(\l)=\Delta(\bar{\l})$ and $\Delta(\l)$ is real for real $\l$.

The eigenvalues of the monodromy matrix  are  called  the Floquet multipliers.
They are the roots of the quadratic equation 
\bay\label{qe}
 w^2 -2\Delta w +1=0,
\ey
and given by the formula 
\bay\label{fqe}
w=\Delta +  \sqrt{\Delta^2-1}.
\ey
The values of $\l: w(\l)=\pm 1$ constitute the points of the periodic/antiperiodic 
spectrum. This condition is equivalent to $\Delta(\l)=\pm1$.
The self--adjointness of the Dirac operator \ref{dirc} implies that points of the spectra are real. 
The NLS hierarchy preserves the periodic/antiperiodic spectrum.

\begin{exa}{\bf The monodromy matrix for the trivial potential $\psi\equiv 0$.} \end{exa} \noi
Let $\psi \equiv 0$. The corresponding transition  matrix  can be easily computed
$M(x,y,\l)=e^{-{i\lt}\sigma_3 (x-y)}$. 
We have $\Delta(\l)=\cos \l l$ and double
eigenvalues at the points $\l_n^{\pm}=\frac{\pi n}{  l}.$
If n is even/odd, then the corresponding $\l_n^{\pm}$ belongs to the  periodic/
anti-periodic spectrum.

The Floquet multipliers  become single--valued on the spectral curve
$$
\Gamma=\{Q=(\l,w)\in { \C}^2:\quad   R(\l,w)=\det\[M(-l,\l)-wI\]=0\}.
$$
The plane  curve   consists of a two sheets covering the plane of the spectral parameter $\l$. For a generic periodic potential from the phase space $\MM$ the 
genus of this curve if infinite, \cite{MV}. To describe geometry of $\Gamma$  we consider  the class of finite gap potentials. 
We assume that there are  a finite number, namely $g+1$,   open  gaps in the spectrum 
$$
\hdots<\l_{n-1}^{-}=\l_{n-1}^{+}<\l_{n}^{-}<\l_{n}^{+}<\hdots< \l_{n+g}^{-}<\l_{n+g}^+< \l_{n+g+1}^{-}=\l_{n+g+1}^{+}<\hdots
$$
Figure 4 below presents an exampe of the discriminant $\Delta(\l)$ 
for a 3 gap potential.

\begin{figure}[ht]
\includegraphics[width=0.80\textwidth]{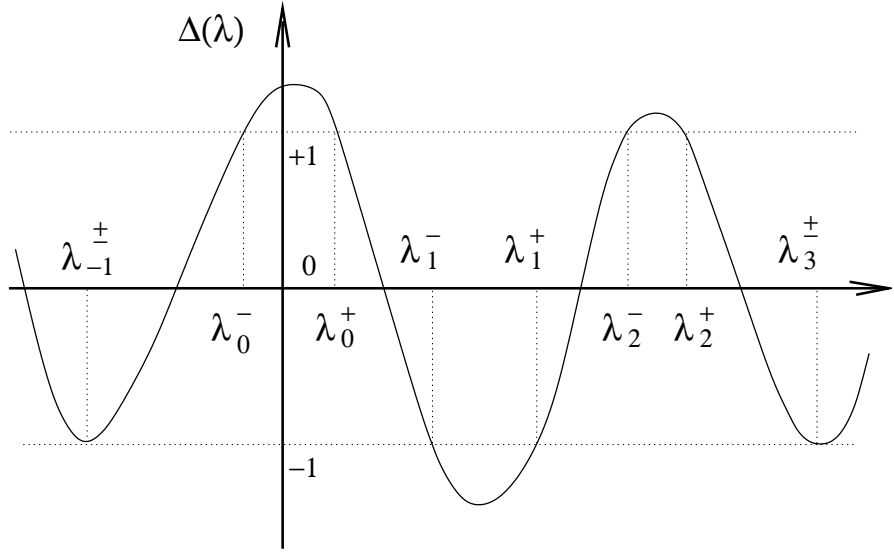}
\caption{The discriminant $\Delta(\l)$ 
for 3 gap potential .}
\end{figure}

\begin{exa}\label{tfm} {\bf The Riemann surface for the trivial potential $\psi\equiv 0$.} \end{exa} \noi
We have $\Delta(\l)=\cos \l l$ and the quadratic equation \ref{qe} has the  solutions $w(\l)=e^{\pm il \l}$. 
The Riemann surface $\G=\G_+ + \G_-$ is reducible and consists of two copies of $\CP^1$ which intersect each other at the points of the double spectrum   $\l_n^{\pm}$.  Each part  $\G_+$ or $\G_-$ has the corresponding infinity $P_+$ or $P_-$. 
The Floquet multipliers are  single valued on $\G$: 
\bey
w(Q)&=&e^{+i\l l},\qquad\qquad\qquad\qquad Q\in \G_+;\label{tasf}\\
w(Q)&=&e^{-i\l l},\qquad\qquad\qquad\qquad Q\in \G_-.\label{tassf}
\eey

For a finite--gap potential the Riemann surface $\G$ is irreducible. 
The spectral  curve is biholomorphicaly  equivalent to the  hyperelliptic curve with branch points at the points of the simple spectrum 
$$
\hat{\Gamma}=\{\hat{Q}=(\l,y)\in { \C}^2:\quad   \hat{R}(\l,y)=y^2+\prod\limits_{k=n}^{n+g} (\l_k^--\l)(\l_k^+-\l)=0\}.
$$
To establish the correspondence  we need   Hadamard product, see \cite{MV}:
$$
\D^2(\l)-1=-\prod\limits_{\Z} \frac{(\l_k^+-\l)(\l_k^--\l)}{a_k^2},
$$
where  $a_0=\frac{1}{l}$ and $a_k=\frac{\pi k}{l}    ,\; k \neq 0$. 
The $1$ to $1$ map $$\hat{Q}=(\l,y) \longrightarrow Q=(\l,w)$$ is defined by the formula
$$
w=\D(\l)+\prod\limits_{\Z-\{n,\hdots,n+g\}} \frac{(\l_k^\pm-\l)}{a_k}\times\prod\limits_{\{n,\hdots,n+g\}} \frac{1}{a_k}\times  y, 
$$ 
which follows from \ref{fqe} and the Hadamard product. 
The Riemann surface $\hat{\G}$ is the desingularization of the spectral curve $\G$.

There are three possible types of important points on $\Gh$. These are the singular points, the points above $\l=\infty$ and the branch points which we now
discuss  in detail.  
\begin{itemize}
\item The singular points are determined by the condition
$$
\partial_{\l}\hat{R}(\l,y)=\partial_{y}\hat{R}(\l,y)=0.
$$
There are no singular points on $\Gh$. 
\item There are two nonsingular points $\Ph_+$ and $\Ph_-$ above $\l=\infty$. At these points\footnote{The notation $\Qh\in (\Ph)$ means that the point $\Qh$ is in the vicinity 
of the point $\Ph$.}  
\bay
w(\Qh)&=&e^{+i\l l}\(1+O\({1/\l}\)\),\qquad\qquad \Qh\in (\Ph_+);\label{asf}\\
w(\Qh)&=&e^{-i\l l}\(1+O\({1/\l}\)\),\qquad\qquad \Qh\in (\Ph_-).\label{assf}
\ey
\item The branch points are specified by the condition
$$
\partial_{y}\hat{R}(\l,y)=0.
$$
They are different from the singular points and correspond to the simple  periodic/antiperiodic spectrum. We denote these points by  
$\hat{\lab}_k^{\pm}=(\l^{\pm}_k,0),$ $\;k=n,\hdots,n+g$. There are $2(g+1)$ of them, each of which  
has a  ramification index 1. 
\end{itemize}
\noi
The Riemann-Hurwitz formula implies that the genus of the curve $\Gh$ is equal to $g$, one less than the number of open gaps in the spectrum.

The map between $\Gh$ and $\G$ transforms  the infinities $\Ph_{-}$ and $\Ph_{+}$ into corresponding punctures $P_{-}$ and $P_{+}$.
It also maps the points $\hat{\lab}_k^{\pm}$ to $\lab_k^{\pm}=(\l^{\pm}_k,(-1)^k)$.
\begin{rem}\label{singu}  Contrary to the case of hyperelliptic surface $\Gh$, there are  singular points on the plane curve $\G$. 
These are the points $(\l^{\pm}_k, (-1)^k)$ of the double spectrum. At these points two sheets of the curve $\G$ intersect. These points accumulate at 
infinities $P_{\pm}$. Asymptotically they form  an arithmetic sequence and approach infinities from {\it real} directions.  
\end{rem}

 Let  $\epm$ be   a 
holomorphic involution on the curve $\Gamma$ permuting sheets 
$$\epm: \; (\l,w)\longrightarrow (\l,1/w).$$ 
The fixed points of $\epm$ are the branch points of $\G$. The involution $\epm$ permutes infinities $\epm: P_-\longrightarrow P_+$.
Let us also  define on $\Gamma$  an antiholomorphic involution  
$$\epsilon_a: (\l, w ) \longrightarrow (\bar{\l}, \bar{w}).$$ The involution $\ea$ also permutes infinities and commutes with $\epm$.
Points of the curve above  open gaps $\[\l_k^-, \l_k^+\]$  form   fixed ``real'' ovals of $\epsilon_a$.   
There are $g+1$ real ovals $a_1,\hdots , a_{g+1}$.

Since $w(Q)$ never vanishes on $\Gamma$  we can define the   quasimomentum $p(Q)$ by the formula $w(Q)=e^{i p(Q)2l}$.  
 Evidently,  $p(Q)$  is defined up to $\frac{\pi n}{ l}$, where $n$ is an integer. 
The asymptotic expansion for $p(Q)$ at infinities can be easily computed
$$
 p(Q)= \pm \lt + p_{0}^\pm \mp  \frac{p_{1}}{ \l} \mp \frac{p_{2}}{ \l^2}\ldots,
\quad \quad \quad \l=\l(Q),\quad Q\in(P_{\pm}),
$$
where 
\beq\label{pinol}
p_{0}^\pm  = \frac{\pi k_{\pm}}{ l}, \;\qquad\qquad k_{\pm} \quad\text{is an integer,}
\eeq
and 
$$
p_{1}  =   \frac{1}{ l} \HH_1, \quad
p_{2}  =   \frac{1}{ l}  \HH_2, \quad
p_{3}  =   \frac{1}{ l} \HH_3, \quad etc.  
$$
Moreover,  the function $w(Q)+ w(\epsilon_\pm Q)$ does not depend on the sheet and it is equal to $2\Delta(\l)$.  
Thus $\Delta(\l(Q))=\cosh i p(Q) 2l$ and the formula 
$$
dp= \pm \frac{1}{ i 2l} d \cosh^{-1} \Delta(\l) = \pm \frac{1}{ i 2l}\frac{d \Delta(\l)}{  \sqrt{\Delta^2-1}}
$$
implies that the  differential $dp$ is of the second kind with double poles at the infinities:
$\pm dp=d\(\lt +O(1)\)$.  The  formula also implies that the differential $dp$ is pure complex on the real ovals. At the same time, the condition  $w(\lab_k^-)=w(\lab_k^+)$ requires   the increment $p(\lab_k^+)-p(\lab_k^-)$ to be real.   
 Therefore,  
$$
\int_{a_k} dp=0,\;\qquad\qquad\qquad \quad k=1,\hdots,g+1.
$$
Since the Floquet multiplies are single--valued on $\G$, for the $b$--periods we have 
\beq\label{pc}
\int_{b_k} dp=\frac{\pi n_{b_k}}{ l},\;\qquad\qquad\qquad n_{b_k} \in \Z,\;\quad k=1,\hdots,g.
\eeq
These are   the {\it periodicity conditions}, \cite{NMP}.  This completes our description of the curve $\G$ for  finite gap potentials. 
All this can be extended with evident modifications to the general infinite genus case. 

Now we consider general smooth potentials from $\MM$ and define the 2-vector 
$$
\ee(x,Q)=\[\begin{array}{ccccc}
\ee_1(x,Q)\\
\ee_2(x,Q)\end{array} \],\qquad\qquad\qquad Q\in \G,
$$
to be an eigenvector of the monodromy matrix $M(x,Q)$ corresponding to the eigenvalue $w(Q)$:
$$
M(x,Q)\ee(x,Q)=w(Q)\ee(x,Q).
$$
Evidently, the components of the vector $\ee$ are defined up to a multiplicative constant, but their ratio is defined uniquely.   
Let us introduce the Weyl function $\P(x,Q)$ on the Riemann surface $\Gamma$ by the formula
\beq\label{weyl}
\P(x,Q)=\frac{\ee_2(x,Q)}{\ee_1(x,Q)}.
\eeq  

We constructed for each $x$ the  direct spectral map from the space  $\MM$ of all smooth  periodic potentials to  the space of pairs: 
\beq\label{dstnls}
\MM\longrightarrow (\G,\P(x,Q)).
\eeq
The direct spectral transform \ref{dstnls}
is invertible. It follows from the general discussion in \cite{V1}. 
Therefore, for each $x$ there exists  an inverse spectral transform
\beq\label{istnls}
(\G,\P(x,Q)) \longrightarrow \MM.
\eeq

In order to establish analytic properties of the function $\P(x,Q)$ we relate it to the Floquet solutions of the spectral problem \ref{spro}.  
The Floquet solution is a  vector--function
$$
\e(x,y,Q)= \[\begin{array}{ccccc}
e_1(x,y,Q)\\
e_2(x,y,Q)\end{array} \],
$$
which is the solution of the auxiliary spectral problem \ref{spro}  with the property
\beq\label{mono}
\e(x+2l,y, Q) =  w(Q) \e(x,y, Q)
\eeq
for all $x$.   It is easy to check that if  this identity  holds at some point $x_0$ 
then it holds for all $x$. This condition  determines the Floquet solution up to a multiplicative constant. The variable $y$ plays the role of a parameter.  The Floquet solution is  uniquely determined by the normalization   condition 
\bay\label{norm}
e_1(y,y,Q) +e_2(y,y,Q)=1.
\ey
Evidently, 
\beq\label{weylf}
\P(x,Q)=\frac{e_2(x,y,Q)}{e_1(x,y,Q)}
\eeq
and the Weyl function  is $2l$--periodic $\P(x+2l,Q)=\P(x,Q)$. 

\begin{exa}{\bf The Floquet solution for the trivial potential $\psi\equiv 0$.} \end{exa} \noi  The Floquet solution is
given by the formula
$$
\e(x,y,Q) = e^{+i \lt(x-y)} \e_0 = e^{+i \lt(x-y)} 
\[\begin{array}{ccccc}  0\\ 1\end{array}\], \qquad\qquad Q\in \G_+,
$$
$$ 
\e(x,y,Q) = e^{-i \lt(x-y)} \hat \e_0 = e^{-i \lt(x-y)} \[\begin{array}{ccccc}
1\\
0\end{array} \],   \qquad\qquad Q\in \G_-.
$$
It has  no poles in the affine part of the curve. We put $\P(x,Q)=\infty$ when $Q\in \G_+$ and $\P(x,Q)=0$ when $Q\in \G_-$.

\begin{exa}\label{topze} {\bf  The one gap  potential $\psi(x)=C e^{i\pi xn/l}$.} \end{exa} \noi  
 In this case the only one gap 
$\l_n^-< \l_n^+$ is open and all other are  closed. We consider the desingularized curve 
$$
\hat{\Gamma}=\{\hat{Q}=(\l,y)\in { \C}^2:\quad   y^2=- (\l_n^--\l)(\l_n^+-\l)\}.
$$ 
Let
$$
\alpha =\frac{\l_n^++\l_n^-}{2},\qquad\qquad\qquad \eta=\frac{\l_n^+-\l_n^-}{2}.
$$
The equation for $\hat{\G}$ in the  new notation can be written as $y^2=\eta^2-(\alpha-\l)^2$. 
According to Lemma 2, \cite{V0}, we have 
$$
\alpha=\frac{\pi n}{l},\qquad\qquad
\eta=2|C|.
$$
The rational curve has the  representation 
\bey
\l&=& z+\frac{\eta^2}{4(z-\alpha)},   \\
y&=& i(z-\alpha)+\frac{\eta^2}{4i(z-\alpha)}, 
\eey
with the uniformization parameter $z=z[\Qh]$. Evidently $z[\Ph_+]=\infty$ and $z[\Ph_-]=\alpha$.  In the following we identify the points 
on the curve and their coordinate $z=z(\Qh)$. 

Let  $\epm$ be   a 
holomorphic involution on the curve $\hat{\Gamma}$ permuting sheets: 
$$\epm: \; (\l,y)\longrightarrow (\l,-y).$$ 
Denoting  $\epm z =z(\epm \Qh)$ we have 
$$
\epm z=  \alpha +\frac{\eta^2}{4(z-\alpha)}.
$$
Let us also  define on $\hat{\Gamma}$  an antiholomorphic involution  
$$
\epsilon_a: (\l, y ) \longrightarrow (\bar{\l}, \bar{y}).
$$
For this antiinvolution we have 
$$
\ea z=  \alpha+\frac{\eta^2}{4(\overline{z}-\alpha)}.
$$ 
Points on the   fixed ``real'' oval are points of the form
$$
z=\alpha +\frac{\eta}{2} e^{i\varphi},\qquad\qquad\qquad \varphi \in \RB^1. 
$$  

The function $p(\Qh)=p(z)$ is single valued  and is given by the formula
$$
p(z)=\frac{1}{2}\(z-\frac{\eta^2}{4(z-\alpha)}\).
$$
It is normalized in a such a way that in the expansion at infinity, $p_0^+=0$ and $p_0^-=\alpha$. It can be verified directly, 
that 
$$
p(z)+p(\epm z)=\alpha. 
$$

In order to write the Floquet solution let us introduce the image of the pole $\gh=\gh(y)$:
$$
z[\gh]= \alpha + z_{\gh}=\alpha +\frac{\eta}{2} e^{i\varphi_{\gh}}.
$$
Evidently, it lies on the real oval. 
We also introduce  two functions 
$$
h_{+}(z|\gh)=\frac{z-\alpha}{z-\alpha-z_{\gh}}   ,\qquad\qquad\qquad h_{-}(z|\gh)=-\frac{z_{\gh}}{z-\alpha -z_{\gh}}. 
$$
Evidently,
$$
h_{+}(z|\gh)+h_{-}(z|\gh)=1.
$$
It  can be verified directly
$$
h_{+}(\epm z|\gh)=h_{-}(z|\epm \gh),
$$
and 
$$
h_{-}(\epm z|\gh)=h_{+}(z|\epm \gh).
$$

Now we can write an explicit formula for the Floquet solution 
$$
\e(x,y,z)=\e(x,y,\Qh)=\[\begin{array}{cccc}
h_-(z|\gh(y))&e^{-i\alpha(x-y)}\\
h_+(z|\gh(y))& \end{array} \]e^{ip(z)(x-y)}. 
$$
Finally, take $x=y$ in the formula 
and use the   formula for the Floquet solution 
$$
\X(x,z)=\frac{e_2(x,x,z)}{e_1(x,x,z)}=\frac{h_+(x,z)}{h_-(x,z)}=-\frac{z-\alpha}{z_{\gh(x)}}.
$$
We see that $\P(x,\Qh)$ has a fixed pole at $z=\infty$ which corresponds to $\Ph_+$. It also has a fixed zero at $z=\alpha$ which corresponds to 
$\Ph_-$.  
This observation completes our considerations of the rational case.

For a general finite gap potential the situation is more complicated. We will show that the 
Floquet solution is given by the explicit formulas \ref{floq}-\ref{coef}. 
Now we  will  state the properties of the Floquet solutions.
\begin{lem}
\label{floc}
 The Floquet solution satisfies the identity
\bay\label{aif}
e(x,y,\ea Q)=\sigma_1 \overline{\e(x,y,Q)}.
\ey
The Floquet solution $\e(x,y,Q)$ has poles common for both components at the points 
$$\gamb_1(y),\gamb_2(y),\hdots,\gamb_{g+1}(y).$$ 
The poles depend on the normalization point $y$. 
The projections of poles $\mu_k(y)= \l(\gamb_k(y))$ are  real. Each $\gamb_k(y)$ lies on the real oval above the corresponding open
gap $[\l_k^-,\l^+_k]$. 
The  first component  $e_1(x,Q)$ has  $g+1$   zeros  at the points
$$
\sigb_1(x),\; \sigb_2(x),\hdots,\sigb_g(x),\sigb_{g+1}=P_+.
$$ 
The  second component  $e_2(x,Q)$ has  $g+1$   zeros at the points
$$
\sigb_1'(x),\; \sigb_2'(x),\hdots,\sigb_g'(x),\sigb_{g+1}'=P_-.
$$ 
For both components the first $g$  zeros depend on the parameter $x$.
In the vicinity of infinities  the function $\e(x,y,Q)$ has the asymptotic behavior 
$$
\e(x,y,Q) =  e^{\pm i \lt(x-y)} \[ \e_0 / \hat \e_0 + o(1)\], \qquad \qquad
Q\in (P_{\pm}).
$$
\end{lem}

\noi
{\it Proof.}\footnote{The complete proof of the Lemma with $y=-l$  can be found in \cite{V4}.}
The proof  is based on the explicit formula for the  Floquet solution $\e(x,y,Q)$:  
\bay\label{floq}
\e(x,y,Q)=A(y,Q) \[\begin{array}{cccc}
m_{11}\\
m_{21} \end{array} \](x,y,\l) +   (1-A(y,Q)) \[\begin{array}{cccccc}
m_{12}\\
m_{22} \end{array} \](x,y,\l),
\ey
where $\l= \l(Q)$  and  the first component  $A(y,Q)$ is 
\bay\label{coef}
A(y,Q)=\frac{M_{12}}{ M_{12}-M_{11} +w(Q)}(y,\l)=\frac{w(Q)- M_{22}}{ M_{21}- M_{22} +w(Q)}(y,\l).
\ey
\qed

The Floquet solution $\e(x,y,Q)$ near  infinities can be expanded into the asymptotic series 
$$
\e(x,y,Q)=e^{+i \lt(x-y)}  \sum\limits_{s=0}^{\infty} \e_s(x,y) \l^{-s} =
e^{+i \lt(x-y)} \sum\limits_{s=0}^{\infty} \[\begin{array}{ccccccc}
b_s \\
d_s
\end{array} \] \l^{-s}, \quad   Q\in (P_{+}),
$$
$$
\e(x,y,Q)=e^{-i \lt(x-y)} \sum\limits_{s=0}^{\infty} \hat \e_s(x,y) \l^{-s}
= e^{-i \lt(x-y)} \sum\limits_{s=0}^{\infty} \[\begin{array}{ccccc}
\overline{d}_s \\
\overline{b}_s
\end{array} \] \l^{-s} ,  \quad Q\in (P_{-}),
$$
with   
$$
b_0= 0,\quad\qquad\qquad d_0=1,
$$
and
$$
b_1=-i\overline{\psi}(x), \qquad\qquad\qquad
d_1=i\overline{\psi}(y) - i \int\limits_{y}^{x} |\psi|^2 dx'.
$$
Since $\ea$ permutes infinities the formula \ref{aif} implies that the asymptotic expansions  are connected.

Now take $x=y$ in    formula \ref{weylf} and use formulas \ref{floq}--\ref{coef}. For the Weyl function we have
\bay\label{def}
\P(x,Q)= \frac{1-A(x,Q)}{A(x,Q)}=\frac{w(Q)-M_{11}(x,\l)}{M_{12}(x,\l)}=\frac{M_{21}(x,\l)}{w(Q)-M_{22}(x,\l)}.
\ey
Lemma \ref{floc}  implies the following   analytic   properties of the Weyl function $\P(x,Q)$ for  finite gap potentials  
\begin{itemize}
\item
 For a fixed $x$ the function $\P(x,Q)$ is meromorphic on $\Gamma$ and satisfies the identity 
\beq\label{inv}
\P(x,\ea Q)=\frac{1}{\overline{\P(x,Q)}}.
\eeq
This implies that on the real ovals $\P$ takes values on the unit circle.   
\item 
It  has $g$ poles $\sigb_k(x),\;k=1,\hdots,g,$ and   zeros  $\sigb_k'(x),\;k=1,\hdots,g$. These poles and zeros depend on the parameter $x$. 

\item The function $\P$  has a simple pole at  $\sigb_{g+1}=P_+$ and a zero at $\sigb_{g+1}'=P_-$. 
In the vicinity of  these points  it has  the asymptotic expansion
\bay
\P(x,Q)&=&\frac{\l}{-i\overline{\psi}(x)}+O(1),\qquad\qquad\qquad\qquad\qquad Q\in(P_+);\label{ass}\\
\P(x,Q)&=&\frac{i\psi(x)}{\l}+O\(\frac{1}{\l^2}\),\qquad\qquad\qquad\qquad\quad Q\in(P_-).\label{asss}
\ey 
\end{itemize}
\noi
Properties \ref{inv} and \ref{ass}--\ref{asss} are true for all potentials from the phase space $\MM$.

\begin{rem}
In the finite gap case the proof of injectivity for the map \ref{dstnls} can be obtained by pure algebraic methods. Indeed  when the function $\P(x,Q)$ is known for some fixed value of $x$, then   
two  sets of zeros $\sigb_k(x),\;k=1,\hdots,g+1,$ and   $\sigb_k'(x),\;k=1,\hdots,g+1,$ are known for each  component
   of the Floquet solution. The components are the Baker--Akhiezer function  on $\G$, \cite{Kri}.  They are  determined uniquely by the  standard asymptotic at infinities and the set of poles or zeros.  Thus the pair $(\G,\P(x,Q))$  determines  the  Baker-Akhiezer function\footnote{$\bullet$ here signifies the argument of the function.}  $\e(\bullet,x,Q)$ and correspondingly  the potential $\psi(\bullet)$. 

\end{rem}

\begin{rem}
We constructed for each $x$ the  direct spectral map from the space  of $g+1$ gap potentials to  the space of pairs   $ (\G,\P(x,Q)).$
The target space has dimension $2g+2$. Indeed 
the spectral curve $\G$ is specified by $2g+2$ real branch points. The periodicity conditions \ref{pc} cut $g$ real degrees of freedom. Also on such a  curve there exists a distinguished differential $dp(Q)$ of the second kind. The differential $dp(Q)$ determines the multivalued function 
$p(Q)$. The coefficient 
$p_0^\pm$ of its asymptotic expansion has the form \ref{pinol} and determines the branch points up to a discrete set of shifts. Therefore, the 
spectral curve $\G$ is specified by  $g+1$ real parameters.  
The  function $\e(\bullet,x,Q)$ is uniquely determined by the fixed asymptotic at infinities  and the poles $\gamb_k(x)$. These poles parametrize the set of all meromorphic functions $\P(x,Q)$. Therefore the set of functions $\P(x,Q)$ for fixed $x$ is topologicaly equivalent to $g+1$ dimensional real torus formed by the real ovals $a_1\times \hdots \times a_{g+1}$
(see Figure 5).
\begin{figure}[ht]
\includegraphics[width=0.80\textwidth]{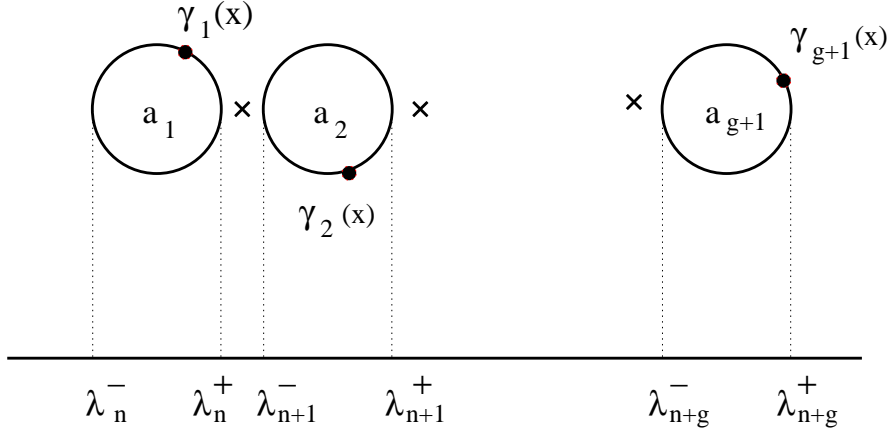}
\caption{The  $g+1$ dimensional torus of potentials.}
\end{figure}

The space of all smooth periodic potentials is stratified manifold. Each strata has an even dimension (possibly infinite). 
\end{rem}

\begin{rem} In example \ref{topze} we considered the simplest case of a rational curve. In this case the only zero and pole stay fixed and do not depend on the parameter $x$ at all. 

For $g >0$ when $x$ changes over a period, the pole $\sigb_k(x),\;k=1,\hdots,g,$ makes a closed loop on the 
surface $\G$. We denote this loop by $s_k=\{\sigb_k(x), x\in \RB^1\}$. The zeros $\sigb_k'(x),\;k=1,\hdots,g,$ of  
the function $\P(x,Q)$ have the same property. 
We denote the corresponding loops by $s_k'=\{\sigb_k'(x), x\in \RB^1\}$. Due to \ref{inv} we have  $\ea \sigb_k(x)=\sigb_k'(x)$. 
The topology of the set of zeros  is an interesting question which we do not address here.  
\end{rem}

\subsection{\bf Analytic properties of the Floquet multipliers.} In this section we explain  the relation between the Floquet solutions and more general, so called, Weyl solutions. 

Let  $\f^T$ denote a transposition of vector $\f$ and let $\f^*$ denote the  adjoint of the vector $\f$; $\LLL^2(a,b)$ be a space of vector functions with the property
$$
\int\limits_{a}^{b} \f^*(x,\l)\f(x,\l) \,dx < \infty.
$$
For a periodic potential, the  
quadratic equation \ref{qe} implies that for each value $\l_0$ of the spectral parameter  we have  $w(Q)w(\epm Q))=1$, where $\l(Q)=\l_0$.  Therefore, either   
$$
|w(Q))|  < 1,\qquad |w(\epm Q))|>1;
$$
or 
$$
|w(Q))| = |w(\epm Q))|=1.
$$
The set of all $\l\in \C$, where the first/second possibility occurs is called the instability/stability area. For an unstable $\l$ the Floquet 
solution  corresponding to the sheet where $|w(Q))|  < 1$ grows  exponentially when $x\rightarrow  - \infty$, while the Floquet 
solution on another sheet  with $|w(\epm Q))|>1$ grows  exponentially when $x\rightarrow  + \infty$. For a stable $\l$  both Floquet solutions stay bounded for all values of $x$.  
Stability  of solutions of periodic (in the $x$-variable) systems is a classical subject, see \cite{K, GL}. We need just some  
elementary facts of this theory. 

If the potential vanishes identically (see Example \ref{tfm}), then all $\l$ with $\Im \l \neq 0$ form the  instability area. The real $\l$ belong to the stability area. For a finite gap potential  all $\l$ with $\Im \l \neq 0$ remain unstable, but all open gaps on the real line belong  to the instability area (Figure 6). 

\begin{figure}[ht]
\includegraphics[width=0.80\textwidth]{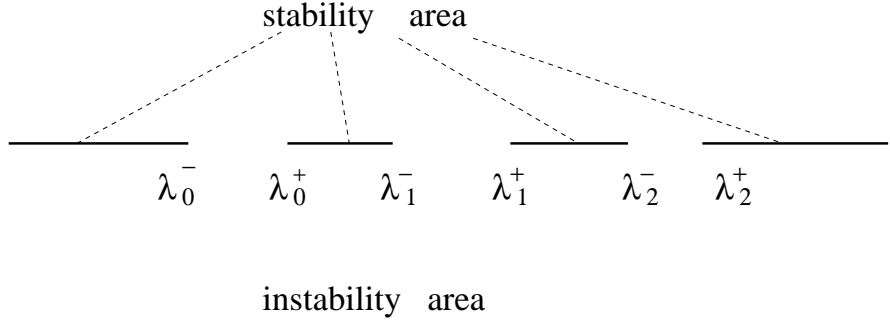}
\caption{Stability areas on the $\l$-plane  
for 3 gap potential .}
\end{figure}

The Weyl  solution $\f(x,\l)$ of the  eigenvalue problem \ref{spro} with arbitrary\footnote{not necessarily periodic.} potential
and $\l$  such that $\Im \l \neq 0$ is defined by the property that it  belongs to $\LLL^2[0,+\infty)$ or $\LLL^2(-\infty,0]$, \cite{W}. Evidently, any Weyl solution is defined up to a multiplicative constant. For the Dirac operator such solutions exist and are unique 
under  the sole assumption of continuity of the potential, \cite{LS}.   For a finite-gap potential,  if 
$Q\in \G$ is such that $\Im \l(Q)  \neq 0$, then $|w(Q)| <1$ or $|w(Q)| >1$ and the Floquet solution $\e(x,y,Q)$ belongs to $\LLL^2[0,+\infty)$ or $\LLL^2(-\infty,0]$ correspondingly. 
Therefore, the Floquet solution is a particular case of the Weyl solution.   

In the case of a periodic potential, the hyperelliptic curve $\Gh$ can be obtained by compactification of the spectral cover, supporting the Weyl solutions, see \cite{V1}.  Indeed in this case 
one can take two copies of $\CP^1$ and cut them along the open gaps. Then identifying edges, one obtains the Riemann surface of  genus $g$ equal to one less than the number of open gaps. 
The $2g+2$ points of the simple spectrum become fixed points of the holomorphic involution permuting sheets of the curve. Thus the compactified cover is the hyperelliptic surface $\Gh$. 
   In the previous section we defined this surface by an  algebraic equation. 
   
 \begin{rem}  The characterization of all {\it finite gap potentilas} of the Dirac operator was obtained along these lines in the paper \cite{dCJ}. In their arguments de Concini and Jonhson    analytically extend the functions originally defined on the spectral cover to the functions on  smooth hyperelliptic curve. 
\end{rem}
\subsection{ The linear fractional transformations  of the Weyl function.}
Let $\f(x,\l)$ be  some Weyl solution of the auxiliary spectral problem \ref{spro} corresponding to some fixed value of the spectral parameter.  
For any $x $  the function $\f(x,\l)$ is  a vector from $\C^2$. It can be represented in the form
$$
\f(x,\l)=c^1(x) \kk_1 + c^2(x) \kk_2,
$$
where 
$$\kk_1=\[\begin{array}{ccccc}
1 \\
0
\end{array} \],\qquad \qquad\qquad \kk_2= \[\begin{array}{ccccc}
0 \\
1
\end{array} \]. 
$$
The quantities $c^1(x)$ and $c^2(x)$ are defined up to a multiplicative constant, but their ratio
$$
\P(x)=\frac{c^2(x)}{c^1(x)}
$$
is defined uniquely. 
Consider $\tilde{\kk}_1,\tilde{\kk}_2$,  some other bases of $\C^2$. Then 
$$
\f(x,\l)=\tilde{c}^1(x) \tilde{\kk}_1 + \tilde{c}^2(x) \tilde{\kk}_2,
$$
and the function $\tilde{\P}(x)$ can be  defined as before:
$$
\tilde{\P}(x)=\frac{\tilde{c}^2(x)}{\tilde{c}^1(x)}.
$$
If the two bases are connected by the formula
\bey
\kk_1&=&a \tilde{\kk}_1 +b \tilde{\kk}_2,\\
\kk_2&=& c \tilde{\kk}_1 +d \tilde{\kk}_2;
\eey
where $a,b,c$ and $d$ are complex constants, then the function $\tilde{\P}(x)$ and $\P(x)$ are connected by the linear  fractional transformation 
\beq\label{lft}
\tilde{\P}=\frac{d \P+b}{c\P+a}.
\eeq

Our definition of the Weyl function $\P(x,Q)$ in   formula \ref{weylf} corresponds to the  bases $\kk_1,\kk_2$ in $\C^2$.
Any other  bases $\tilde{\kk}_1,\tilde{\kk}_2$ will lead to a new Weyl type function $\tilde{\P}(x,Q)$ connected to $\P(x,Q)$ by  formula \ref{lft}. 

Geometrically the transformation \ref{lft}  changes the  global uniformization parameter in the target space for the map
$$ 
\P(x,Q): \G\longrightarrow \CP^1.
$$

\section{The Poisson bracket.}
\subsection{ The Poisson bracket for the Weyl function.} In this section we compute the image of the Poisson bracket 
\ref{nlspb}  under the direct spectral transform \ref{dstnls}.

Let us introduce the function  
$$
\O(Q)=\frac{w^2(Q)+1}{w^2(Q)-1}=\coth ip(Q)2l.
$$
It is easy to see that
$$
\O(\epm Q)=-\O(Q),
$$
and 
$$
\O(\ea Q)=\overline{\O}(Q).
$$
In the finite part of the curve there are  two types of important points. These are singular points and branch points. 
The function $\l(Q)$ where $Q=(\l,w) \in \G$ is a local parameter everywhere on the curve with  the exception of these  points. 
At the branch and singular points the function $\O(Q)$ has  
poles because  $w^2(Q)= 1$. The zeros of the function coincide with the points of the anti-periodic spectrum of the problem 
on  the doubled interval $2\times 2l$.  

Now we are ready to state the main result of this section. 

\begin{thm}
\label{comp}
 Let $Q=(\l,w(Q))$ and $P=(\m,w(P))$ be  different from the branch and singular points  of $\Gamma$ and poles of $\X$. The bracket \ref{nlspb} for two functions $\P(Q)=\P(x,Q)$ and $\P(P)=\P(y,P)$ 
is given by the formula
\bay\label{mah}
\{\P(Q),\P(P)\}=-2\times \frac{\(\P(Q)-\P(P)\)^2}{\l-\mu}\times \frac{\O(Q)+\O(P)}{2}.
\ey 
\end{thm}

First for the proof we need some identity for the quartic products of solutions of the 
auxiliary spectral problem.

\begin{lem}\label{ident} 
Let the column vectors $\gb^+(x,\m),\; \f^+(x,\l)$ satisfy 
$$
{\gb^+}'(x,\m)=V(x,\m)\gb^+(x,\m), \qquad\qquad\qquad {\f^+}'(x,\l)=V(x,\l)\f^+(x,\l);
$$
and the row vectors $\gb^-(x,\m),\; \f^-(x,\l)$ satisfy
$$
{\gb^-}'(x,\m)=-\gb^-(x,\m)V(x,\m), \qquad\qquad\qquad {\f^-}'(x,\l)=-\f^-(x,\l)V(x,\l).
$$ 
The following identity holds:
\beq\label{idt}
f_2^+f_1^-g_1^+g_2^- -f_1^+f_2^-g_2^+g_1^-=\frac{1}{i(\l-\m)}\frac{d }{d x} \[(\f^-I\gb^+)(\gb^-I\f^+)\].
\eeq
\end{lem} 
\noi
{\it Proof.} The identity can be verified by differentiation. \qed

The next lemma provides  the   bracket 
for  entries of  the transition matrix. 

\begin{lem}\label{rpb} The following formulas hold for the entries $m_{ij}(\l)=m_{ij}(x,y,\l),$ $ x>y,$ of the transition matrix $M(x,y,\l)$
\bay
\{m_{11}(\l),m_{12}(\m)\}=K\times \frac{m_{12}(\l)m_{11}(\m)-m_{12}(\m)m_{11}(\l)}{\l-\m}\\
\{m_{11}(\l),m_{21}(\m)\}=K\times \frac{m_{11}(\l)m_{21}(\m)-m_{11}(\m)m_{21}(\l)}{\l-\m}\\
\{m_{11}(\l),m_{22}(\m)\}=K\times \frac{m_{12}(\l)m_{21}(\m)-m_{12}(\m)m_{21}(\l)}{\l-\m}\\
\{m_{12}(\l),m_{21}(\m)\}=K\times \frac{m_{11}(\l)m_{22}(\m)-m_{11}(\m)m_{22}(\l)}{\l-\m}\\
\{m_{12}(\l),m_{22}(\m)\}=K\times \frac{m_{12}(\l)m_{22}(\m)-m_{12}(\m)m_{22}(\l)}{\l-\m}\\
\{m_{21}(\l),m_{22}(\m)\}=K\times \frac{m_{22}(\l)m_{21}(\m)-m_{22}(\m)m_{21}(\l)}{\l-\m}.
\ey
with $K=-2$.
All other brackets vanish. 
\end{lem}

\noi
{\it Proof.}  We will prove  the first identity.  The other formulas can be proved along the same lines. 

Let $M^{\bullet}=\d M$ be a variation of $M(x,y,\l)$ in response to the variation 
of $\psi(z)$ and $\psib(z),\; y\leq z\leq x$. Then 
$M^{{\bullet}'}=VM^{\bullet} + V^{\bullet}M$. The solution of this 
nonhomogenious equation is  
$$
M^{\bullet}(x,y,\l)=M(x,y,\l)\int\limits_{0}^{x} M^{-1}(\xi,y,\l)V^{\bullet}(\xi) 
M(\xi,y,\l)\,d\xi
$$
Therefore
$$
\frac{\d M(x,y,\l)}{ \d \psi(z)}=M(x,z,\l)\left(\begin{matrix} 0& 0\\
                                                            1& 0 \end{matrix} \right) M(z,y,\l),
$$
$$
\frac{\d M(x,y,\l)}{ \d \psib(z)}=M(x,z,\l)\left(\begin{matrix} 0& 1\\
                                                             0& 0 \end{matrix}
\right) M(z,y,\l). 
$$

Using the formulas for the gradients, we have 
\bey
\{m_{11}(\l),m_{12}(\m)\}=2i\int_{y}^{x} dz\, [m_{11}(x,z,\l)m_{21}(z,y,\l)m_{21}(x,z,\m)m_{12}(z,y,\m)- \\
                                               - m_{12}(x,z,\l)m_{11}(z,y,\l)m_{11}(x,z,\m)m_{22}(z,y,\m)].
\eey 
Since the matrix $M(x,y,\l)$ satisfies the differential equations 
$$
\frac{\partial M(x,y,\l)}{\partial x}= V(x,\l)  M(x,y,\l),\qquad\quad \frac{\partial M(x,y,\l)}{\partial y}= -M(x,y,\l)V(y,\l),
$$
we can apply the identity of Lemma \ref{ident} to the computation of the integral. 
If we put 
$$
\f^+= \[\begin{array}{ccccc}
m_{11}\\
m_{21}
\end{array} \](z,y,\l),   \qquad \qquad \qquad \gb^+=\[\begin{array}{ccccc}
m_{12}\\
m_{22}
\end{array} \](z,y,\m);
$$
and 
$$
\f^-= [m_{11},m_{12}](x,z,\l), \qquad \qquad \gb^-=[m_{11},m_{12}](x,z,\m);
$$
then we can apply the identity in the variable $z$:
$$
\{m_{11}(\l),m_{12}(\m)\}=\frac{2}{\l-\m} \[(\f^-I\gb^+)(\gb^-I\f^+)\]|_{y}^{x}.
$$
Using boundary conditions we obtain  the result. \qed

Now we are ready to prove Theorem \ref{comp}.

\noi
{\it Proof. }  For the function $\P(x,Q)$ we use representation \ref{def}.  We use the result of Lemma \ref{rpb} with  an unspecified value of the constant $K$.  We split  the computation of the Poisson bracket

\bay
\{\P(Q),\P(P)\}&=&\{\frac{w(Q)-M_{11}(\l)}{M_{12}(\l)},\frac{w(P)-M_{11}(\m)}{M_{12}(\m)}\}\nonumber\\ 
&=&\{w(Q)-M_{11}(\l),w(P)-M_{11}(\m)\}\frac{1}{M_{12}(\l)}\frac{1}{M_{12}(\m)}\qquad \label{A}\\
&+&\{w(Q)-M_{11}(\l),\frac{1}{M_{12}(\m)}\}\frac{1}{M_{12}(\l)}(w(P)-M_{11}(\m))\qquad \label{B}\\
&+&\{\frac{1}{M_{12}(\l)},w(P)-M_{11}(\m)\}(w(Q)-M_{11}(\l))\frac{1}{M_{12}(\m)}\qquad \label{C}\\
&+&\{\frac{1}{M_{12}(\l)},\frac{1}{M_{12}(\m)}\}(w(Q)-M_{11}(\l))(w(P)-M_{11}(\m)).\nonumber
\ey
into small steps. 
First note that due to Lemma \ref{rpb} the last line vanishes. 

\noi
{\it Step 1.}  Using the identities 
$$
\{w(Q),M_{11}(\m)\}=K\times \frac{1}{2}\frac{\partial w}{\partial \D}(Q)\frac{M_{21}(\l)M_{12}(\m)-M_{21}(\m)M_{12}(\l)}{\l-\m},
$$
$$
\{w(P),M_{11}(\l)\}=K\times \frac{1}{2}\frac{\partial w}{\partial \D}(P)\frac{M_{21}(\l)M_{12}(\m)-M_{21}(\m)M_{12}(\l)}{\l-\m};
$$
which follow from Lemma \ref{rpb} for \ref{A}, we have 
\bey\label{first}
\{w(Q)&-&M_{11}(\l),w(P)-M_{11}(\m)\}\frac{1}{M_{12}(\l)}\frac{1}{M_{12}(\m)}=\\
=&\phantom{o}&\frac{K}{\l-\m}\times\frac{1}{M_{12}(\l)M_{12}(\m)}\frac{M_{21}(\l)M_{12}(\m)-M_{21}(\m)M_{12}(\l)}{2}\(\frac{\partial w}{\partial \D}(P)-
\frac{\partial w}{\partial \D}(Q)\).
\eey
Introducing the function $\Pi(\l)=\frac{1}{2}(M_{11}(\l)-M_{22}(\l))$ we have 
$$
M_{22}(\l)=\Delta(\l)-\Pi(\l). 
$$
This together with formula   \ref{def}  imply 
$$
M_{21}(\l)=\P(Q)(w(Q)+\Pi(\l)-\D(\l)).
$$
After simple transformations,  \ref{A} becomes
\bay
K\times \frac{1}{(\l-\m)}\left[\frac{1}{M_{12}(\l)}\P(Q)\frac{w(Q)+\Pi(\l)-\D(\l)}{2}\(\frac{\partial w}{\partial \D}(P)-
\frac{\partial w}{\partial \D}(Q)\)-\label{fourth}\right.\qquad \\
-\left. \frac{1}{M_{12}(\m)}\P(P)\frac{w(P)+\Pi(\m)-\D(\m)}{2}\(\frac{\partial w}{\partial \D}(P)-
\frac{\partial w}{\partial \D}(Q)\)\].\qquad \label{fifth}
\ey

\noi
{\it Step 2.} The second   line \ref{B} is   equal to 
$$
-\(\{w(Q),M_{12}(\m)\}-\{M_{11}(\l),M_{12}(\m)\}\)\frac{\P(P)}{M_{12}(\l)M_{12}(\m)}.
$$
The third   line \ref{C} is   equal to
$$
\(\{w(P),M_{12}(\l)\}-\{M_{11}(\m),M_{12}(\l)\}\)\frac{\P(Q)}{M_{12}(\l)M_{12}(\m)}.
$$ 
After simple transformations the sum of \ref{B}-\ref{C} becomes
\bay
&=&\frac{1}{M_{12}(\l)M_{12}(\m)}\(\P(Q)\{w(P),M_{12}(\l)\}-\P(P)\{w(Q),M_{12}(\m)\}\)+\label{second}\\
            &+& \frac{1}{M_{12}(\l)M_{12}(\m)}\(\P(P)\{M_{11}(\l),M_{12}(\m)\}-\P(Q)\{M_{11}(\m),M_{12}(\l)\}\).\qquad \label{third}
\ey

\noi
{\it Step 3.} This step consists of simple manipulations with  \ref{second}-\ref{third}.
Using the identities 
$$
\{w(P),M_{12}(\l)\}=K\times-\frac{\partial w}{\partial \D}(P)\frac{\Pi(\l)M_{12}(\m)-\Pi(\m)M_{12}(\l)}{\l-\m},
$$
$$
\{w(Q),M_{12}(\m)\}=K\times-\frac{\partial w}{\partial \D}(Q)\frac{\Pi(\l)M_{12}(\m)-\Pi(\m)M_{12}(\l)}{\l-\m},
$$ 
which follow from Lemma \ref{rpb},  we have for \ref{second}
\bay
K\times\frac{1}{(\l-\m)}\[\frac{1}{M_{12}(\l)}\(\P(P) \frac{\partial w}{\partial \D}(Q)\Pi(\l) - \P(Q) \frac{\partial w}{\partial \D}(P)\Pi(\l)\)+\right.\label{sixth}\\
+ \left. \frac{1}{M_{12}(\m)}\(\P(Q) \frac{\partial w}{\partial \D}(P)\Pi(\m) - \P(P) \frac{\partial w}{\partial \D}(Q)\Pi(\m)\)\]. \label{seventh}
\ey 

For \ref{third} using Lemma \ref{rpb} and 
$$
M_{11}(\l)=\D(\l)+\Pi(\l),
$$
after simple algebra we obtain 
\bay
K\times\frac{1}{(\l-\m)}\[\frac{1}{M_{12}(\l)}(\Delta(\l)+\Pi(\l))(\P(Q)-\P(P))\right.-\label{eight}\\
-\left.\frac{1}{M_{12}(\m)}(\Delta(\m)+\Pi(\m))(\P(Q)-\P(P))\]\label{ninth}.
\ey

\noi
{\it Step 4.} Collecting terms \ref{fourth}, \ref{sixth} and \ref{eight} containing $M_{12}(\l)$ in the denominator we have
\bey
K\times\frac{1}{(\l-\m)}\frac{\P(Q)}{M_{12}(\l)}
 &\times&\[\frac{w(Q)+\Pi(\l)-\D(\l)}{2}\right.\(\frac{\partial w}{\partial \D}(P)-\frac{\partial w}{\partial \D}(Q)\)-\\
&-&\qquad\left.\Pi(\l)\frac{\partial w}{\partial \D}(P)+ \Delta(\l) +\Pi(\l) \]-\\
K\times\frac{1}{(\l-\m)}\frac{\P(P)}{M_{12}(\l)}&\times&\[\Pi(\l)\frac{\partial w}{\partial \D}(Q)- \Delta(\l) -\Pi(\l) \].
\eey

For the first term we have 
\bey
\frac{\P(Q)}{M_{12}(\l)} &\times&
 \[\frac{w(Q)+\Pi(\l)-\D(\l)}{2}\right.\(\frac{\partial w}{\partial \D}(P)-\frac{\partial w}{\partial \D}(Q)\)
-\left.\Pi(\l)\frac{\partial w}{\partial \D}(P)+ \Delta(\l) +\Pi(\l) \]\\
&=&\frac{\P(Q)}{M_{12}(\l)}\times\[\frac{w(Q)-\Pi(\l)-\D(\l)}{2}\frac{\partial w}{\partial \D}(P)-(w(Q)-\Pi(\l)-\D(\l))\]+\\
&+&\frac{\P(Q)}{M_{12}(\l)}\times\[-\frac{w(Q)+\Pi(\l)-\D(\l)}{2}\frac{\partial w}{\partial \D}(Q)+w(Q)\].
\eey
The  derivative  
$
\partial w/\partial \Delta(Q)  
$
can be easily computed from  the quadratic equation \ref{qe}:  
$$
\frac{\partial w}{\partial \D}(Q)=\frac{w(Q)}{w(Q)-\D(\l)}=\frac {2w^2(Q)}{w^2(Q)-1}.
$$
Therefore
\bey
\P^2(Q)\[\frac{1}{2}\frac{\partial w}{\partial \D}(P)-1\]&+&\frac{\P(Q)}{M_{12}(\l)}\frac{\partial w}{\partial \D}(Q)\[-\frac{w(Q)+\Pi(\l)-\D(\l)}{2}+w(Q)
-\D(\l)\]\\
&=&\P^2(Q)\[\frac{1}{2}\frac{\partial w}{\partial \D}(P)-1 \]   +  \frac{\P^2(Q)}{2}\frac{\partial w}{\partial \D}(Q).
\eey
Note
\beq\label{uuu}
\O(Q)=\frac{\partial w}{\partial \D}(Q)-1.
\eeq
Therefore, for the first term we obtain 
\beq\label{fft}
\P^2(Q)\frac{\O(Q)+\O(P)}{2}.
\eeq

For the second term we have 
\bey
\frac{\P(P)}{M_{12}(\l)}&\times&\[\Pi(\l)\frac{\partial w}{\partial \D}(Q)- \Delta(\l) -\Pi(\l) \]=\\
&=&\frac{\P(P)}{M_{12}(\l)}\times\[\Pi(\l)\frac{\partial w}{\partial \D}(Q)- w(Q)\] +\frac{\P(P)}{M_{12}(\l)}\times\[w(Q)- \Delta(\l) -\Pi(\l) \]=\\
&=&-\P(Q)\P(P)\frac{\partial w}{\partial \D}(Q) +\P(Q)\P(P).
\eey
Using \ref{uuu} we have 
\bay\label{frt}
-\P(Q)\P(P)\O(Q).
\ey
Taking the sum of \ref{fft} and \ref{frt} we finally obtain 
\beq\label{las}
K\times\frac{1}{\l-\m}\[ \P^2(Q)\frac{\O(Q)+\O(P)}{2}
-\P(Q)\P(P)\O(Q)\].
\eeq
\noi
{\it Step 5.} Collecting terms \ref{fifth}, \ref{seventh} and \ref{ninth} containing $M_{12}(\m)$ in the denominator we have
\bey
K\times\frac{1}{\l-\m}\frac{\P(P)}{M_{12}(\m)}
&\times& \[\frac{w(P)+\Pi(\m)-\D(\m)}{2}\right.\(\frac{\partial w}{\partial \D}(Q)-\frac{\partial w}{\partial \D}(P)\)-\\
&-&\qquad\left.\Pi(\m)\frac{\partial w}{\partial \D}(Q)+ \Delta(\m) +\Pi(\m)\]-\\
K\times\frac{1}{\l-\m}\frac{\P(Q)}{M_{12}(\m)}&\times&\[\Pi(\m)\frac{\partial w}{\partial \D}(P)- \Delta(\m) -\Pi(\m) \].
\eey
After we interchange $\l \leftrightarrow \m$ and $Q \leftrightarrow  P$  this expression becomes identical to the formula of Step 4. Therefore, it is equal to 
\beq\label{last}
K\times\frac{1}{\l-\m}\[ \P^2(P)\frac{\O(Q)+\O(P)}{2}
-\P(Q)\P(P)\O(P)\]
\eeq
Taking the sum of \ref{las} and \ref{last} we obtain formula  \ref{mah}. 
\qed

\begin{rem} It is interesting to note that the bracket of two meromorphic functions on the Riemann surface is not a meromorphic function in any reasonable sense. The presence of infinitely many intersection or branch points makes it possible to obtain any limiting value for the bracket, 
when one of the points, say $Q$, is fixed and 
$P$ tends to infinity. This is due to the fact that $\P(P)$ viewed as  Hamiltonian can open gaps arbitrarily  far, and produce a change 
of topology  of the curve.\footnote{I indebted to I. Krichever for this remark.} 

\end{rem}

\begin{rem} The formula \ref{mah} formally contains all three cases of  Theorem 4.1 in \cite{V1}. One has to replace $\O$ on $+1$ or $-1$ to  get the formulas obtained there. 
\end{rem}

\begin{rem}\label{asahb} We call the formula \ref{mah} the deformed Atiyah-Hitchin bracket.
When  the points $Q,\,P$ approach infinities     from imaginary directions (see Remark \ref{singu}) then $w(Q)$ and $w(P)$ tend    to infinity or zero. 
Therefore     
$\O(Q)$ and $\O(P)$ tend to $+1$ or $-1$. For example, if $Q,P \in P_+$ and $\l(Q)=i\tau,\;\; \l(P)=i\theta$ where $\tau,\theta \rightarrow +\infty$, then 
$$
\O(Q)\rightarrow -1, \qquad\qquad\qquad \O(P)\rightarrow -1;
$$ 
and formula  \ref{mah} becomes 
$$
\{\P(Q),\P(P)\}\sim + 2 \times \frac{\(\P(Q)-\P(P)\)^2}{\l-\mu}.
$$
 Thus the formula  asymptotically    
coincides with  the bracket for the classical Weyl function  in the open upper/lower hulf-plane, see \cite{V1}. 
\end{rem}

\begin{rem}
The deformed AH bracket is invariant under linear--fractional transformations. Namely, if we introduce the new Weyl type function $\tilde{\P}$ by  formula \ref{lft}, then the bracket for  $\tilde{\P}$ is given by the same formula \ref{mah}. The proof is identical to the pure AH case, see \cite{V1}.
Another consequence of invariance is the formula for the bracket of the first component $A(x,Q)$. Note, first 
$$
A(x,Q)=\frac{1}{1+\P(x,Q)}.
$$
This  implies that $A(x,Q)$ is the meromorphic function with $g+1$ zeros at  $$\sigb_1(x),\hdots,\sigb_g(x), \sigb_{g+1}.$$  
Therefore, for $Q,P$ different from poles of $A$ and also branch and singular points we have 
\beq\label{fftt}
\{A(x,Q),A(x,P)\}=-2\times \frac{\(A(x,Q)-A(x,P)\)^2}{\l(Q)-\l(P)}\times \frac{\O(Q)+\O(P)}{2}.
\eeq
\end{rem}

\subsection{Computation of the Poisson bracket for field variables.} 
The goal of this section is to show that the deformed AH bracket \ref{mah} can be taken as a starting point for the construction of the Poisson formalism. The inverse spectral transform \ref{istnls}  maps the deformed AH bracket   to  the phase space. Therefore, 
the bracket for the  field variables $\psi(x)$ and $\psib(x)$ can be obtained from the formula for the deformed AH bracket. 
Namely, we prove the following theorem  

\begin{thm}\label{ppbb} The deformed AH bracket for the Weyl function implies the follwowing Poisson brackets for the field variables\footnote{The identities are understood 
in the sense of generalized functions: $u(x)=v(x)$ if for any $f(x)\in C^{\infty}_0$ we have $\int u(x) f(x) dx= \int v(x) f(x) dx$. }:
\bay
\{\psi(z),\psi(y)\}&=&0,\label{pp}\\
\{\psib(z),\psib(y)\}&=&0,\label{pop}\\
\{\psib(z),\psi(y)\}&=&2i \d(z-y).\label{thi}
\ey
These identities are an equivalent form of the Poisson bracket \ref{nlspb}. 
\end{thm}
 We use Remarks \ref{singu}, \ref{asahb} and the strategy developed in \cite{V1} for the proof of the analogous result.

Unfortunately, the inverse  spectral transform  \ref{istnls}  is very implicit.   
However, if the function $\P(x,Q)$ is known for all $x$ then  \ref{ass}-\ref{asss} imply  the formulas for the potential:  
\beq\label{psib} 
\lim_{Q \rightarrow P_+} \frac{\l(Q)}{\P(x,Q)}=-i\psib(x)
\eeq
and 
\beq\label{psi}
\lim_{Q \rightarrow P_-} \l(Q)\P(x,Q)=i\psi(x).
\eeq
The limits are complex conjugate of each other due to \ref{inv} and 
\bey
\lim_{Q\rightarrow P_-} \l(Q) \P(x,Q)&=& \lim_{Q\rightarrow P_-} \l(\ea \ea Q) \P(x,\ea \ea Q)=
\lim_{Q\rightarrow P_-} \overline{\frac{\l( \ea Q)} {\P(x, \ea Q)}}\\
&=&\lim_{Q\rightarrow P_+} \overline{\frac{\l(  Q)} {\P(x,  Q)}}.
\eey
Formulas \ref{psib}--\ref{psi} allow  us to  effectively solve the inversion problem
\beq\label{nistnls}
(\G,\P(x,Q),\; x\in \RB^1) \longrightarrow \MM.
\eeq

This approach requires the formula  for   the bracket 
$
\{\P(x,Q),\P(y,P)\},
$
for $x\neq y$. This bracket is computed asymptotically when one of the points $Q$ or $P$  approaches infinity from imaginary direction, see Remark \ref{singu}.

\begin{lem}\label{rrt}  Let $Q \rightarrow  P_{\pm}$ and the point $P$ is fixed, then 
\beq\label{asahd}
\{\P(y,Q),\P(x,P)\}\sim  e^{-i\l(Q)(x-y)}\{\P(x,Q),\P(x,P)\}. 
\eeq
\end{lem}
 
\noi
{\it Proof.} The proof is  analogous to the  corresponding result for the pure AH bracket, \cite{V1}.  The identity   
$$
\e(x,y,Q)=M(x,y,\l) \e(y,y,Q),\qquad\qquad\qquad \l=\l(Q), 
$$ 
implies  
$$
\P(x,Q)=\frac{m_{22}(x,y,\l)\P(y,Q)+ m_{21}(x,y,\l)}{m_{12}(x,y,\l)\P(y,Q)+ m_{11}(x,y,\l)}.
$$ 
Note, when $Q \rightarrow  P_{\pm}$,  
$$
M(x,y,\l)\sim e^{-\frac{i\l}{2}\sigb_3(x-y)}.
$$
Therefore,
$$
\P(x,Q)\sim e^{i\l(Q)(x-y)}\P(y,Q).
$$
The asymptotic in not uniform in the $x$-variable.  Thus the left hand side is periodic in $x$ but the expression at the right is not!
\qed
 
The next lemma establishes that the Poisson tensor is real. 
\begin{lem}\label{pbreal} The Poisson brackets for the field variables $\psi(x)$ and $\psib(x)$ are real 
\bey
\overline{\{\psi(y),\psi(z)\}}&=&\{\psib(y),\psib(z)\},\label{firre}\\
\overline{\{\psib(y),\psi(z)\}}&=&\{\psi(y),\psib(z)\}.\label{secre}
\eey
\end{lem}

\noi
{\it Proof.} The proof requires the  previous Lemma. The rest is  analogous to the  corresponding result for the pure AH bracket, \cite{V1}.
\qed

Now we are ready to prove the main result.

\noi
{\it Proof of Theorem \ref{ppbb}.}   Due to Lemma \ref{pbreal},  identities \ref{pp} and \ref{pop} are equivalent. 
We compute  the bracket \ref{pp}. Let $Q,P\in \G_R$. Using  formula \ref{psi} and Lemma \ref{rrt} for  $f(x)\in C^{\infty}_0$ and $y \leq z$ we have: 
\bey
&\phantom{000}&\int\limits_{-\infty}^{z} dy  f(y) \{\psi(z),\psi(y)\}=-\int\limits_{-\infty}^{z} dy f(y) \{i\psi(z),i \psi(y)\}\\
&=&-\lim_{Q,P\rightarrow P_-} \int\limits_{-\infty}^{z} dy f(y) \{\l(Q)\P(z,Q),\l(P)\P(y,P)\}\\
&=& -\lim_{Q,P\rightarrow P_-}  \l(Q)\l(P)\int\limits_{-\infty}^{z} dy f(y) \{\P(z,Q),\P(y,P)\}\\
&=& -\lim_{Q,P\rightarrow P_-}  \l(Q)\l(P) \{\P(z,Q),\P(z,P)\}\int\limits_{-\infty}^{z} dy f(y)e^{-i\l(P)(z-y)}.\\
\eey
Let $Q,P$ be such that $\l(Q)=-i\tau,\;\l(P)=-2i\tau $.
Since $\P$ has a zero at $P_-$ we have using Theorem \ref{comp} when $\tau \rightarrow +\infty$, 
\bey
\lim_{Q,P\rightarrow P_-}  \l(Q)\l(P) \{\P(z,Q),\P(z,P)\}\sim \l(Q)\l(P) {(\P(z,Q)-\P(z ,P))^2\over \l(Q) - \l(P)}=O(\tau^{-1}). 
\eey
For the integral   we have 
\bey
\int\limits_{-\infty}^{z} dy f(y)e^{-i\l(P)(z-y)}  =O(\tau^{-1}).
\eey
Therefore,
$$
\{\psi(z),\psi(y)\}=0,\qquad\qquad\qquad\qquad y\leq z.
$$
Using skew symmetry of the bracket and interchanging $y$ and $z$ we obtain 
$$
\{\psi(z),\psi(y)\}=0,\qquad\qquad\qquad\qquad y\geq z.
$$
Taking the sum of these two formulas,  we obtain \ref{pp}.

Now we compute the bracket \ref{thi}. Let  $Q\rightarrow P_+,\; P\rightarrow P_-$. Then  using formulas \ref{psib}, \ref{psi} and Lemma \ref{rrt} for  $f(x)\in C^{\infty}_0$ and $y \leq z$ we have: 
\bey
&\phantom{000}&\int\limits_{-\infty}^{z} dy  f(y) \{\psib(z),\psi(y)\}=\int\limits_{-\infty}^{z} dy f(y) \{-i\psib(z),i \psi(y)\}\\
&=&\lim\int\limits_{-\infty}^{z} dy f(y) \{\frac{\l(Q)}{\P(z,Q)},\l(P)\P(y,P)\}\\
&=& \lim -\frac{\l(Q)\l(P)}{\P^2(z,Q)}\int\limits_{-\infty}^{z} dy f(y) \{\P(z,Q),\P(y,P)\}\\
&=& \lim -\frac{\l(Q)\l(P)}{\P^2(z,Q)}\{\P(z,Q),\P(z,P)\}\int\limits_{-\infty}^{z} dy f(y) e^{-i\l(P)(z-y)}.
\eey
Using Theorem \ref{comp}, 
\bey
\{\P(z,Q),\P(z,P)\}= -2\times \frac{\(\P(Q)-\P(P)\)^2}{\l(Q)-\l(P)}\times \frac{\O(Q)+\O(P)}{2}.
\eey
Since $\P$ has a pole at $P_+$ and a zero at $P_-$,  we have asymptotically
\bey
-\frac{\l(Q)\l(P)}{\P^2(z,Q)}&&\{\P(z,Q),\P(z,P)\}\sim\\
 &&\sim -\frac{\l(Q)\l(P)}{\P^2(z,Q)}\times -2 \times \frac{\P^2(z,Q)}{\l(Q)-\l(P)} \times \frac{\O(Q)+\O(P)}{2} .
\eey
Let $P=\ea Q$ and  $\l(Q)=i\tau,\; \tau \rightarrow +\infty$; then 
$$
\O(Q)\rightarrow -1,\qquad\qquad\qquad \O(P)\rightarrow -1;
$$
and 
\bey
-\frac{\l(Q)\l(P)}{\P^2(z,Q)}&&\{\P(z,Q),\P(z,P)\}\sim -\frac{2\l(Q)\l(P)}{\l(Q)-\l(P)}.
\eey
Using   steepest decent we have 
\bey
\therefore&=& \lim_{\tau \rightarrow +\infty} -\frac{2\tau^2}{i\tau +i\tau}\int\limits_{-\infty}^{z} dy f(y)  e^{-\tau(z-y)}=if(z).
\eey
Therefore,
\beq\label{alo}
\{\psib(z),\psi(y)\}=i \d(z-y),\qquad\qquad\qquad\qquad y\leq z.
\eeq
Using the realness of the bracket, by Lemma   \ref{pbreal} we have 
$$
\{\psi(z),\psib(y)\}=-i \d(z-y),\qquad\qquad\qquad\qquad y\leq z.
$$
By the skew symmetry of the bracket, interchanging $z$ and $y$, 
\beq\label{aloo}
\{\psib(z),\psi(y)\}=i \d(z-y),\qquad\qquad\qquad\qquad z\leq y.
\eeq
Taking the sum of \ref{alo} and \ref{aloo}, we obtain  \ref{thi}.

\qed

\subsection{Functions of the first component.}
The moving poles of the function $\P$ do not seem to be an appropriate object for the construction of canonical coordinates. Indeed there are $g$ complex poles but $g+1$ real degrees of freedom  in the isospectral set of potentials. The trouble is that the function $\P$ does not carry explicitly the information about the 
real poles of the Baker--Akhiezer functions. Both components have the same poles and this information disappears when we divide one component of the Baker--Akhiezer functions by another. 
What will happen if one will take another meromorphic function of the first component? Some natural choices are considered in this section.

The Floquet solution $\e(x,y,Q)$ satisfies the identity
$$
\[J\partial_x- J V\] \e(x,y,Q) =0, \qquad\qquad\qquad J=i\sigb_2; 
$$
which is just another way to write an auxiliary  spectral problem \ref{spro}.
Let us define the dual Floquet  solution $\e^+(x,y,Q)=\[e_{1}^+(x,y,Q), e_{2}^+(x,y,Q)\]$ at the point
$Q$ by 
$$
\e^+(x,y,Q)=\e(x,y,\epsilon_{\pm }Q )^T.
$$
The dual  Floquet solution 
$\e^+(x,y,Q)$ satisfies\footnote{The action of the differential operator 
$D=\sum\limits_{j=0}^{k} \omega_{j}
\partial^{j}$ on the row vector $f^{+}$ is defined as $f^+ D=\sum\limits_{j=0}^{k} (-\partial)^{j}(f^{+} \omega_j).$}
$$
\e^+(x,y,Q)\[J\partial_x-J V\]=0.
$$ 
Introduce the Wronskian 
function $\F(y,Q)$ by the formula
$$
\F(y,Q)=\e^+(x,y,Q)J \e(x,y,Q).
$$
The fact that  $\F$  does not depend on $x$ can be verified by differentiation.   
Evidently,  
$$
\F(y,Q)=<\e^+(x,y,Q)J\e(x,y,Q)>= \frac{1}{ 2l} \int_{-l}^{l}\e^+(x,y,Q)J\e(x,y,Q) dx.
$$
As it  follows from \ref{floq}--\ref{coef}, 
\bay\label{ndef}
\F(y,Q)= A(y,\epm Q)- A(y,Q).
\ey
Now we list   the properties of $\F(y,Q)$ which follow from Lemma \ref{floc} for finite gap potentials. 
\begin{itemize}
\item The function $\F(y,Q)$ is meromorphic on $\G$ and satisfies the identity 
\bey
\F(y,\epm Q)&=&-\F(y,Q),\\
\F(y,\ea Q)&=&-\overline{\F(y,Q)}.
\eey

\item The function $\F(y,Q)$ has  $2(g+1)$ poles, on both sheets at $\gamb_k(y)$ and $\epm \gamb_k(y)$ on the real ovals,  and $2(g+1)$ 
zeros at the branch points $\lab_k^{\pm}$.

\item At infinities  $\F(y,Q)$ has the asymptotic behavior 
\begin{eqnarray*}
 \F(y,Q) & =& +1 + \frac{i\psib(y)-i\psi(y)}{ \l} + O\(\frac{1}{ \l^2}\), 
\qquad \qquad \quad \quad
Q\in (P_+), \\
\F(y,Q) & =&- 1 - \frac{i\psib(y)-i\psi(y)}{ \l} + O\(\frac{1}{ \l^2}\), 
\qquad \qquad \quad \quad
Q\in (P_-). 
\end{eqnarray*}
\end{itemize}

Let us  introduce another function 
$$
\Xi(y,Q)=A(y,\epm Q)+ A(y,Q).
$$
In  the next theorem the bracket for  $\F(x,Q)$ will be expressed  with the  help of the  function $\Xi(y,Q)$ and the traditional  $\O(Q)$ and  $\l(Q)$. 
This result  will be used for  construction of canonical variables. 
 
\begin{thm}\label{popd} Suppose $Q=(\l,w(Q))$ and $P=(\m,w(P))$ are not the branch or crossing points  of $\Gamma$; they are not poles of the functions $\F(Q)=\F(x,Q)$ and $\Xi(P)=\Xi(x,P)$. Then the bracket \ref{nlspb} for the  functions $\F(Q)$ and $\Xi(P)$ 
is given by the formulas
\bey\label{mahn}
\{\F(Q),\F(P)\}=-2\times \frac{(\Xi(Q)-\Xi(P))(\F(P)\O(Q)-\F(Q)\O(P))}{\l-\m},\\
\{\Xi(Q),\Xi(P)\}=-2\times \frac{(\Xi(Q)-\Xi(P))(\F(P)\O(P)-\F(Q)\O(Q))}{\l-\m}.
\eey
\end{thm}

\noi
{\it Proof.}
We   compute  the bracket  directly using formula \ref{fftt}.  
\qed

\subsection{ Canonical variables.} The present section  relates our results with approach of Novikov, Veselov and Dubrovin, \cite{NV, DN}. 
The next result is well known.
 
\begin{thm}\label{ffm} \cite{FM}. If $\gamb_k=\gamb_k(y),\; k=1,\hdots, g+1,$ and $y\in \RB^1$, then the following identities hold:
\bay
\{\l(\gamb_k),\l(\gamb_n)\}&=&0,\label{fiid}\\
\{p(\gamb_k),p(\gamb_n)\}&=&0,\\
\{p(\gamb_k),\l(\gamb_n)\}&=&\d_n^k.
\ey
\end{thm}

\noi
{\it Proof of Theorem \ref{ffm} (incomplete).} When $Q$ is not a branch or crossing point, then $\l(Q)$ plays the role of a local parameter. 
If $\gamb_k,\;\gamb_n$ are such  points, then  the  function $\F(x,Q)$  in the vicinity of these points can be written as  
$$
\F(x,Q)=\frac{\varphi_k(x,\l(Q))}{\l(Q)-\l(\gamb_k)}, \qquad\qquad\qquad Q\in (\gamb_k);
$$
and 
$$
\F(x,P)=\frac{\varphi_n(x,\l(P))}{\l(P)-\l(\gamb_n)}, \qquad\qquad\qquad P\in (\gamb_n).
$$
Then, by standard properties of the Poisson bracket,  
$$
\{\F(x,Q),\F(x,P)\}=\frac{\{\l(\gamb_k),\l(\gamb_n)\}}{(\l(Q)-\l(\gamb_k))^2(\l(P)-\l(\gamb_n))^2}+\hdots .
$$
Dots signify   terms  of order lower than four when $Q\rightarrow \gamb_k$ and $P\rightarrow \gamb_n$. From another side,  
using the result of Theorem  \ref{popd} we have 
\bey
\{\F(x,Q),\F(x,P)\}=\frac{O(1)}{\l(Q)-\l(\gamb_k)} +\frac{O(1)}{\l(P)-\l(\gamb_n)}.
\eey
The highest order of the pole on  the right is one. 
These imply the first identity \ref{fiid}. 
\qed

Evidently we did not use finite-gap property anywhere in the proof. 
It seems to be an important task to complete the proof of the Theorem using our approach. We will return to this question elsewhere.


\vskip 1in
\noindent
Department of Mathematics
\newline
Michigan State University
\newline
East Lansing, MI 48824
\newline
USA
\vskip 0.3in
\noindent
vaninsky@math.msu.edu

\end{document}